\documentclass[osajnl,twocolumn,showpacs,groupedaddress,10pt]{revtex4-1}
\usepackage{amsmath,amssymb}
\usepackage{graphicx}
\usepackage{pgfplots}
\usepackage{subfigure}
\usepackage{xcolor}
\usepackage{pdfpages}

\begin{document}
\title{Analysis of enhanced stimulated Brillouin scattering \\ in silicon slot waveguides}

\author{Rapha\"{e}l Van Laer, Bart Kuyken, Dries Van Thourhout and Roel Baets}
\affiliation{Photonics Research Group, Ghent University--imec, Belgium \\ Center for Nano- and Biophotonics, Ghent University, Belgium \vspace{-5mm}}
\email{raphael.vanlaer@intec.ugent.be}

\begin{abstract} Stimulated Brillouin scattering has attracted renewed interest with the promise of highly tailorable integration into the silicon photonics platform. However, significant Brillouin amplification in silicon waveguides has yet to be shown. In an effort to engineer a structure with large photon-phonon coupling, we analyzed both forward and backward Brillouin scattering in high-index-contrast silicon slot waveguides. The calculations predict that gradient forces enhance the Brillouin gain in narrow slots. We estimate a currently feasible gain of about $10^{5} \, \text{W}^{-1}\text{m}^{-1}$, which is an order of magnitude larger than in a stand-alone silicon wire. Such efficient coupling could enable a host of Brillouin technologies on a mass-producible silicon chip.
\end{abstract}


\maketitle 
\vspace{-5em}
\section{Introduction}

Stimulated Brillouin scattering (SBS) is a nonlinear process which couples optical to mechanical waves \cite{Chiao1964, Shen1965}. It is a powerful means to control light, with applications ranging from lasing \cite{Stokes1982}, comb generation \cite{Braje2009, Kang2009, Savchenkov2011a} and isolation \cite{Kang2011} to RF-waveform synthesis \cite{Li2013a}, slow/stored light \cite{Okawachi2005, Zhu2007a} and reconfigurable filtering \cite{Zadok2007}. With this in mind, SBS has been explored in a wide variety of systems, such as conventional and photonic crystal fibers \cite{Ippen1972,Shelby1985,Kobyakov2009a,Wang2011a,Beugnot2007}, silica microspheres \cite{Tomes2009a, Bahl2011} and wedge-disks \cite{Li2012b}, calcium fluoride resonators \cite{Grudinin2009} and chalcogenide rib waveguides \cite{Pant2011}. Therefore the prospect of strong SBS in small-core silicon wires \cite{Bogaerts2005} is tantalizing.

Such wires are known for their large Kerr and Raman nonlinearity \cite{Lin2007}. However, Brillouin scattering has so far lagged behind in silicon. The culprit is the silica substrate on which the silicon wires are typically made. It severely decreases both the wires' mechanical flexibility and the phonons' lifetime. Unlike in chalcogenide rib waveguides \cite{Pant2011,Poulton2013}, elastic waves in silicon cannot be guided by internal reflection because sound is faster in silicon than in silica.

A theoretical model by Wang et al. \cite{Rakich2012, Qiu2012} recently predicted that the efficiency of SBS would increase dramatically by removing the substrate. Then the elastic waves are confined to the core because of the large acoustic mismatch between air and silicon, although there is still no internal reflection. The model included not just electrostriction but also radiation pressure, which was traditionally neglected as a driver of Brillouin scattering. Thus electrostriction and radiation pressure interfere in nanoscale waveguides, connecting the fields of Brillouin scattering and optomechanics \cite{Kippenberg2007, Kippenberg2008a, Li2008, VanThourhout2010a}. The validity of the new SBS model has been confirmed by recent observations of SBS in a hybrid silicon nitride-silicon waveguide \cite{Shin2013b}, although the enhancement of SBS in silicon-only photonic wires \cite{Rakich2012, Qiu2012} remains unverified.

\begin{figure}
\centering
\label{fig:a}
\subfigure{\begin{tikzpicture}
   		 \node[anchor=south west,inner sep=0] (image) at (0,0) {\raisebox{20pt}{\includegraphics[width=.4\columnwidth]{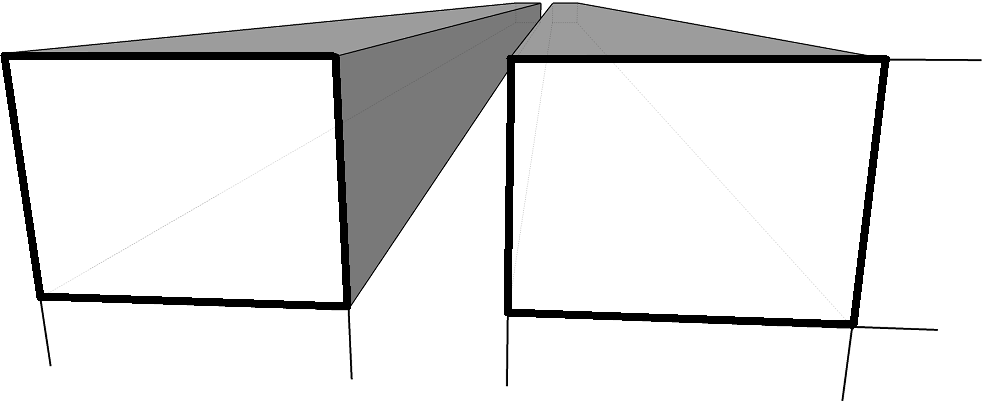}}};
    		\begin{scope}[x={(image.south east)},y={(image.north west)}]
       		\node at (0.1,-0.03) {\textcolor{white}{(a)}};
		\node at (0.2,1.1) {\small{(a)}};
		\node at (0.19,0.4) {\small{$a$}};
		\node at (0.68,0.38) {\small{$\bar{a}$}};
		\node at (0.43,0.38) {\small{$g$}};
		\node at (0.97,0.65) {\small{$b$}};
		\node at (0.07,0.82) {\small{Si}};	
    		\end{scope}
		\end{tikzpicture}
}
\subfigure{\begin{tikzpicture}[scale = 0.4*\columnwidth/(3.5in)]
\pgfplotsset{try min ticks=3}
\pgfplotsset{max space between ticks=50pt}
\begin{axis}[%
width=3.5in,
height=3.13in,
axis on top,
xmin=-0.55,
xmax=0.55,
xlabel={$x$ ($\mu$m)},
ymin=-0.55,
ymax=0.55,
ylabel={$y$ ($\mu$m)},
colorbar horizontal,
colormap/jet,
colorbar style={
at={(0.5,1.03)},anchor=south,
xticklabel pos=upper, font = \LARGE},
point meta min=0,
point meta max=1,
xtick = {-0.4,0,0.4},
ytick = {-0.4,0,0.4}
]
\addplot graphics [xmin=-0.5709375,xmax=0.566015625,ymin=0.5709375,ymax=-0.566015625] {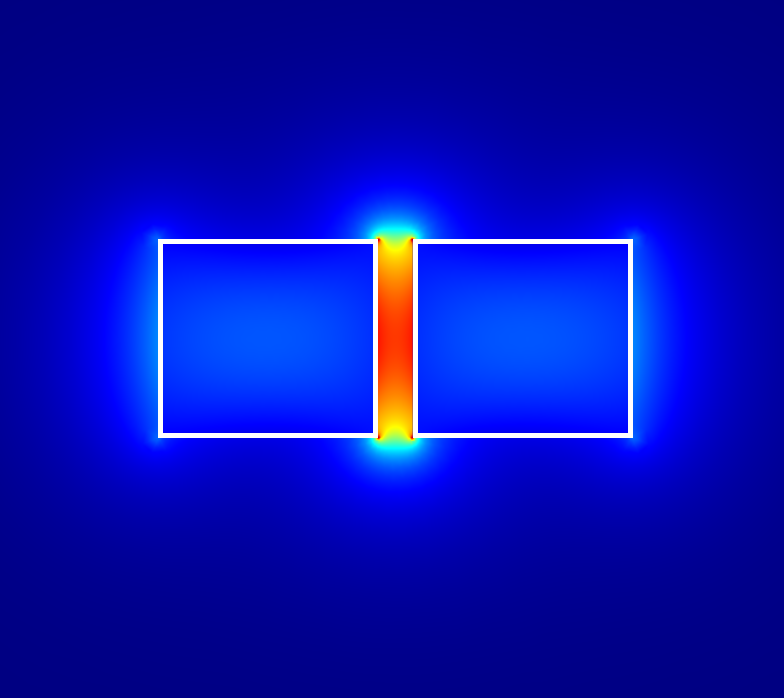};
\tikzstyle{every node}=[font=\LARGE]
\end{axis}
\begin{scope}[x={(image.south east)},y={(image.north west)}]
\node at (0.10,0.99) {\textcolor{white}{\small{(b)}}};
\node at (-0.13,1.12) {\textcolor{white}{\small{(b)}}};
\node[anchor=base] at (0.43,0.04) {\textcolor{white}{$\bf{|E|}^2$}};
\node at (0.58,0.99) {\textcolor{white}{quasi-TE}};
\end{scope}
\end{tikzpicture}%
}
\subfigure{\begin{tikzpicture}
   		 \node[anchor=south west,inner sep=0] (image) at (0,0) {\raisebox{9pt}{\includegraphics[width=.4\columnwidth]{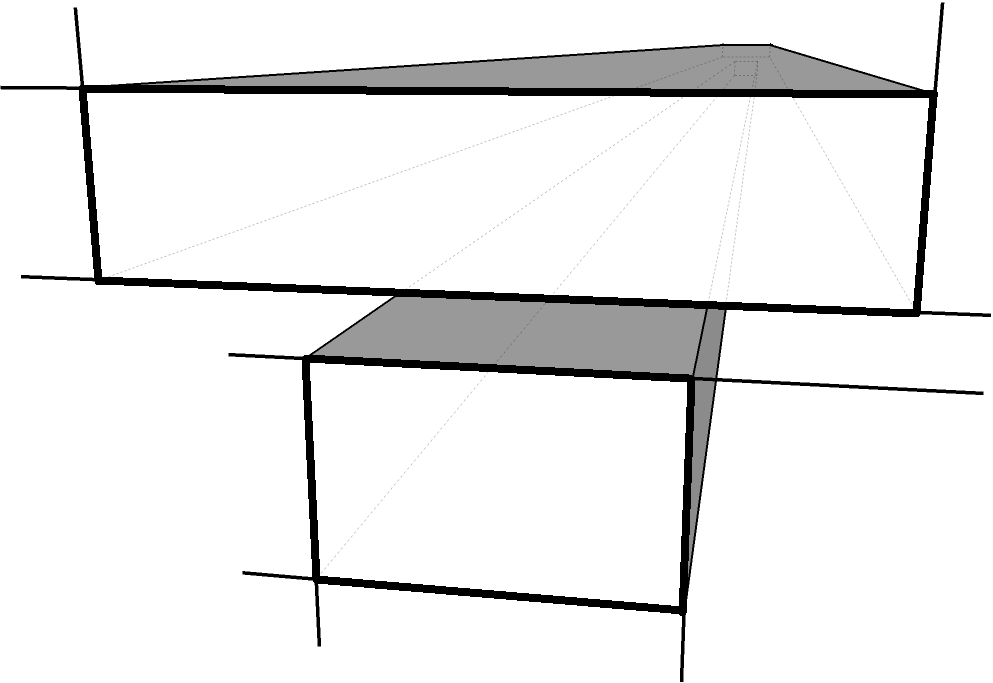}}};
    		\begin{scope}[x={(image.south east)},y={(image.north west)}]
       		\node at (0.1,0.11) {\textcolor{white}{\small{(a)}}};
		\node at (0.16,1) {\small{(c)}};
		\node at (0.24,0.4) {\small{$\bar{a}$}};
		\node at (0.03,0.76) {\small{$a$}};
		\node at (0.97,0.54) {\small{$g$}};
		\node at (0.52,0.98) {\small{$b$}};
		\node at (0.48,0.14) {\small{$\bar{b}$}};
		\node at (0.14,0.82) {\small{Si}};	
    		\end{scope}
		\end{tikzpicture}
}
\subfigure{\begin{tikzpicture}[scale = 0.4*\columnwidth/(3.5in)]
\pgfplotsset{try min ticks=3}
\pgfplotsset{max space between ticks=50pt}
\begin{axis}[%
width=3.5in,
height=3.13in,
axis on top,
xmin=-0.55,
xmax=0.55,
xlabel={$x$ ($\mu$m)},
ymin=-0.55,
ymax=0.55,
ylabel={$y$ ($\mu$m)},
colorbar horizontal,
colormap/jet,
colorbar style={
at={(0.5,1.03)},anchor=south,
xticklabel pos=upper, font = \LARGE},
point meta min=0,
point meta max=1,
xtick = {-0.4,0,0.4},
ytick = {-0.4,0,0.4}
]
\addplot graphics [xmin=-0.5709375,xmax=0.566015625,ymin=0.5709375,ymax=-0.566015625] {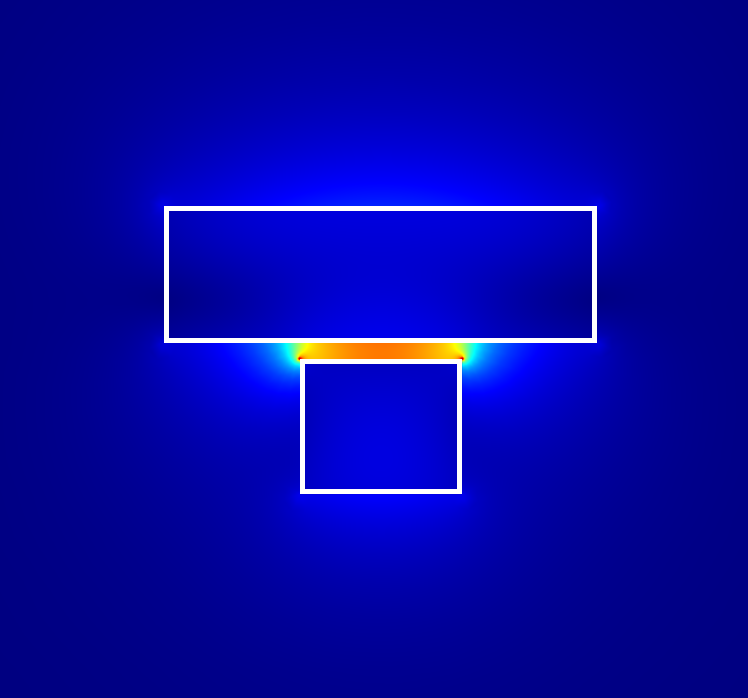};
\tikzstyle{every node}=[font=\LARGE]
\end{axis}
\begin{scope}[x={(image.south east)},y={(image.north west)}]
\node at (0.11,0.78) {\textcolor{white}{\small{(d)}}};
\node[anchor=base] at (0.43,0.04) {\textcolor{white}{$\bf{|E|}^2$}};
\node at (0.59,0.79) {\textcolor{white}{quasi-TM}};
\end{scope}
\end{tikzpicture}%
}
\caption{Vertical (a) and horizontal (c) silicon slot waveguides suspended in air, with the corresponding optical mode (b,d).}
\vspace{-6mm}
\end{figure}
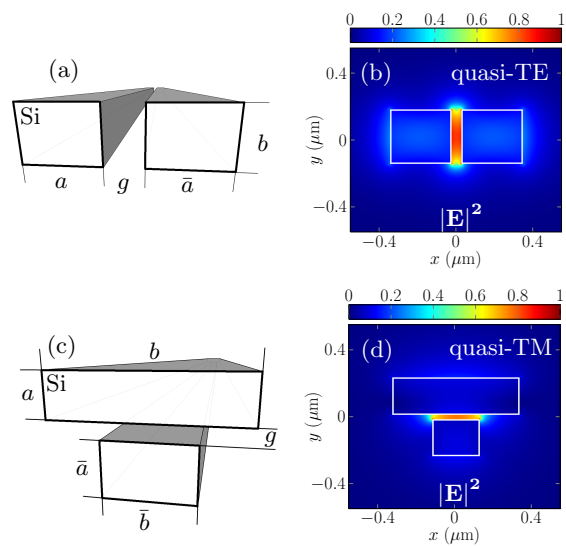

In this Letter we take the study of Brillouin scattering to silicon slot waveguides, to exploit their strong mode confinement \cite{Almeida2004a,Sun2007} and large gradient forces \cite{Li2010}. We perform full-vectorial coupled optical and mechanical simulations of the Brillouin gain coefficient using the finite-element solver {\textsf{\small{COMSOL}}}.

\section{Background and assumptions}

\begin{figure}
\centering
\label{fig:c}
\subfigure{\begin{tikzpicture}
   		 \node[anchor=south west,inner sep=0] (image) at (0,0) {\raisebox{0pt}{\includegraphics[width=.32\columnwidth]{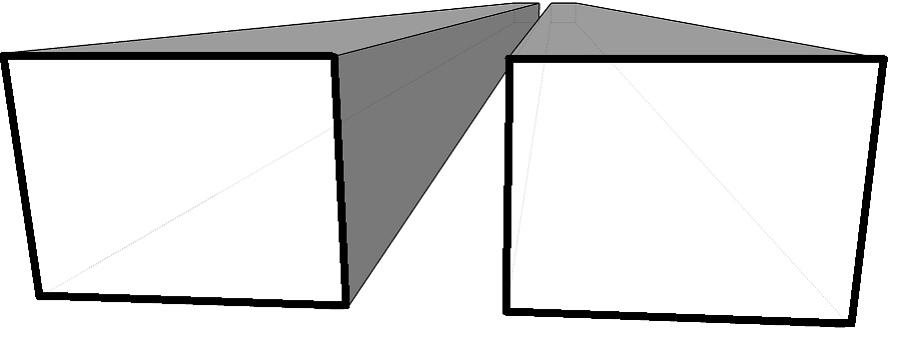}}};
    		\begin{scope}[x={(image.south east)},y={(image.north west)}]
		\node at (0.2,1.15) {\small{(a)}};
		\node at (0.5,1.17) {{$4G$}};
    		\end{scope}
		\end{tikzpicture}
}
\subfigure{\begin{tikzpicture}
   		 \node[anchor=south west,inner sep=0] (image) at (0,0) {\raisebox{0pt}{\includegraphics[width=.32\columnwidth]{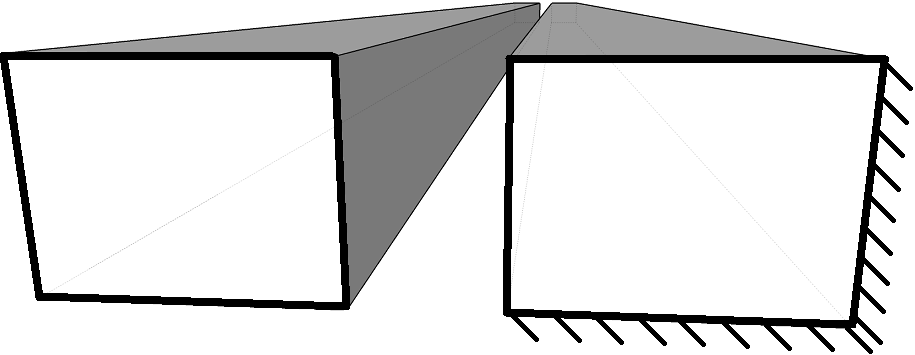}}};
    		\begin{scope}[x={(image.south east)},y={(image.north west)}]
		\node at (0.2,1.15) {\small{(b)}};
		\node at (0.5,1.17) {{$G$}};
    		\end{scope}
		\end{tikzpicture}
}
\subfigure{\begin{tikzpicture}
   		 \node[anchor=south west,inner sep=0] (image) at (0,0) {\raisebox{0pt}{\includegraphics[width=.19\columnwidth]{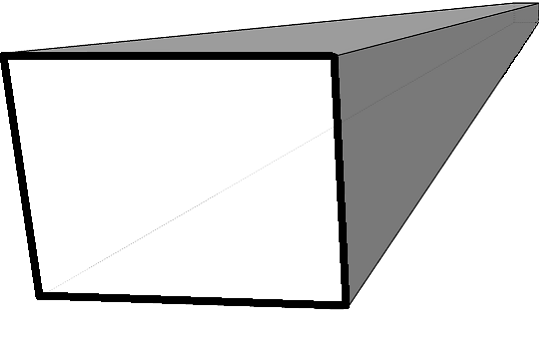}}};
    		\begin{scope}[x={(image.south east)},y={(image.north west)}]
		\node at (0.2,1.15) {\small{(c)}};
		\node at (0.5,1.18) {{$\tilde{G}$}};
    		\end{scope}
		\end{tikzpicture}
}
\vspace{-2mm}
\caption{We compare three scenarios: (a) a slot with two free silicon beams, (b) a slot with one free and one fixed beam and (c) a stand-alone free beam.}
\vspace{-2mm}
\end{figure}
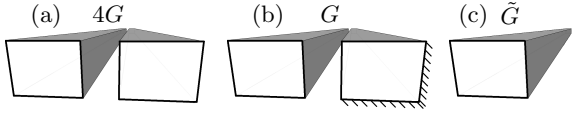

We consider vertical (fig.1a-b) and horizontal (fig.1c-d) slot waveguides suspended in air. Both waveguides strongly confine light, creating large radiation pressure close to the slot.
\begin{figure}
\centering
\label{fig:b}
\subfigure{\begin{tikzpicture}
   		 \node[anchor=south west,inner sep=0] (image) at (0,0) {\raisebox{0pt}{\includegraphics[]{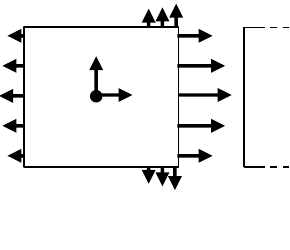}}};
    		\begin{scope}[x={(image.south east)},y={(image.north west)}]
		\node at (0.15,0.98) {\small{(a)}};
		\node at (1.2,0.95) {\textcolor{white}{\small{a}}};
		\node at (0.37,0.98) {$\mathbf{f_{\text{rp}}}$};
		\node at (0.3,0.77) {$y$};
		\node at (0.49,0.61) {$x$};
    		\end{scope}
		\end{tikzpicture}
}
\subfigure{\begin{tikzpicture}
   		 \node[anchor=south west,inner sep=0] (image) at (0,0) {\raisebox{0pt}{\includegraphics[]{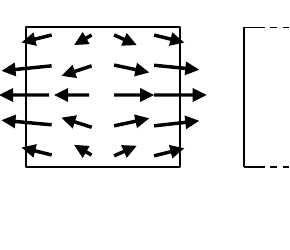}}};
    		\begin{scope}[x={(image.south east)},y={(image.north west)}]
		\node at (0.15,0.98) {\small{(b)}};
		\node at (0.37,0.98) {$\mathbf{f_{\text{es}}}$};
    		\end{scope}
		\end{tikzpicture}
}
%
%
%
\vspace{-10mm}
\caption{Typical optical force profile on left beam of vertical slot waveguide: (a) radiation pressure and (b) electrostrictive body force. The radiation pressure is large close to the slot.}
\vspace{-6mm}
\end{figure}
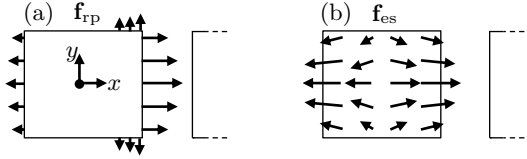
This gives rise to the possibility of (1) improving the photon-phonon coupling, (2) testing SBS theory in a regime dominated by gradient forces and (3) exciting new types of phonons.

If the two silicon beams would be identical (fig.2a), their mechanical resonances could be addressed simultaneously. Then the SBS gain would be $4G$, with $G$ the single-beam gain. However, our simulations show that fabrication imperfections on the order of nanometers are sufficient to shift the phonon spectrum by more than one mechanical linewidth ($\thicksim \hspace{-4pt} 10 \, \text{MHz}$). So we assume just one beam of dimensions $(a,b)$ contributes to SBS (fig.2b). Moreover, we call $\tilde{G}$ the peak gain associated with a phonon in a stand-alone silicon wire (fig.2c).

A particular mechanical mode with displacement $\mathbf{u}$, wavevector $K$, stiffness $k_{\text{eff}}$ and quality factor $Q$ has a peak SBS gain $G$ of $\omega Q|\langle \mathbf{f}, \mathbf{u}\rangle|^{2}/(2k_{\text{eff}})$, with $\omega$ the optical frequency, $\mathbf{f} = \mathbf{f_{\text{rp}}}+\mathbf{f_{\text{es}}}$ the power-normalized optical force distribution and $\langle \mathbf{f}, \mathbf{u} \rangle = \int{\mathbf{f}^{*} \cdot \mathbf{u}}{\, dA}$ the photon-phonon overlap \cite{Rakich2012,Qiu2012}. The radiation pressure $\mathbf{f_{\text{rp}}}$ is located on the waveguide boundaries (fig.3a), while the electrostrictive force $\mathbf{f_{\text{es}}}$ has both a body (fig.3b) and a boundary (not shown) component. The boundary component of $\mathbf{f_{\text{es}}}$ is an order of magnitude smaller than $\mathbf{f_{\text{rp}}}$. Furthermore, we define $G_{\text{rp}}$ and $G_{\text{es}}$ as the SBS gain when only $\mathbf{f_{\text{rp}}}$ or $\mathbf{f_{\text{es}}}$ is present. The total gain $G$ is determined by interference between $\mathbf{f_{\text{rp}}}$ and $\mathbf{f_{\text{es}}}$.

In forward (backward) SBS, the Stokes and pump wave co- (counter-) propagate. Phase-matching then requires that $K \approx 0$ ($K \approx 2\beta$), with $\beta$ the pump wavevector. We launch the Stokes and pump waves into the same mode, leaving inter-modal SBS \cite{Kang2011} for further study. In addition, we work at $\lambda = 1.55 \, \mu\text{m}$ and use a flat $Q$ of $10^{3}$ as in \cite{Rakich2012, Qiu2012}.

\section{SBS in vertical slot waveguides}
Figures 4a-c show the forward and backward SBS spectrum for a vertical slot waveguide with dimensions $(a,b,\bar{a},g)=(315 \, \text{nm},0.9a,a, 50 \, \text{nm})$, including only the three modes with largest gain.

\begin{figure}
\centering
\label{fig:d}
\subfigure{\begin{tikzpicture}[scale = 0.44*\columnwidth/(3.5in)]
\pgfplotsset{try min ticks=3}
\pgfplotsset{max space between ticks=50pt}
\begin{axis}[width=3.5in,
height=2.33in,
xmin=11,
xmax=19,
xlabel={Frequency (GHz)},
ymin=0,
ymax=4.5,
ylabel={Gain ($\text{W}^{-1}\text{m}^{-1}$)},
legend style={fill=none,draw=none,legend cell align=left,at={(0.5,-0.41)},anchor=north, legend columns=-1},
xtick = {10,12,14,16,18,20}
]
\addplot [
color=black,
solid,
line width=2.0pt,
forget plot
]
table[row sep=crcr]{
12.5613262162514 0\\
12.5613262162514 4.17615661806709\\
};
\addplot [
color=black,
solid,
line width=2.0pt,
forget plot
]
table[row sep=crcr]{
18.1720805338954 0\\
18.1720805338954 1.14459439972653\\
};
\addplot [
color=black,
solid,
line width=2.0pt,
forget plot
]
table[row sep=crcr]{
14.8678514134985 0\\
14.8678514134985 0.185345201667898\\
};
\addplot [
color=black,
solid,
line width=2.0pt,
forget plot
]
table[row sep=crcr]{
24.6213482923394 0\\
24.6213482923394 0.148443690211904\\
};
\addplot [
color=black,
solid,
line width=2.0pt,
forget plot
]
table[row sep=crcr]{
29.2954070058747 0\\
29.2954070058747 0.120749528280508\\
};
\addplot [
color=black,
solid,
line width=2.0pt,
forget plot
]
table[row sep=crcr]{
0 0\\
0 0\\
};
\addplot [
color=black,
solid,
line width=2.0pt,
forget plot
]
table[row sep=crcr]{
0 0\\
0 0\\
};
\addplot [
color=black,
solid,
line width=2.0pt,
forget plot
]
table[row sep=crcr]{
0 0\\
0 0\\
};
\addplot [
color=black,
solid,
line width=2.0pt,
forget plot
]
table[row sep=crcr]{
0 0\\
0 0\\
};
\addplot [
color=black,
solid,
line width=2.0pt,
forget plot
]
table[row sep=crcr]{
0 0\\
0 0\\
};
\addplot [
color=blue,
line width=3.0pt,
mark size=5.0pt,
only marks,
mark=x,
mark options={solid}
]
table[row sep=crcr]{
0 0\\
0 0\\
0 0\\
0 0\\
0 0\\
0 0\\
0 0\\
0 0\\
12.5613262162514 0.866335269007811\\
0 0\\
0 0\\
14.8678514134985 0.00969252067286596\\
0 0\\
0 0\\
0 0\\
18.1720805338954 0.0442874840724149\\
0 0\\
0 0\\
0 0\\
0 0\\
0 0\\
0 0\\
24.6213482923394 0.00160967632822347\\
0 0\\
0 0\\
0 0\\
0 0\\
0 0\\
0 0\\
29.2954070058747 0.0123861060438769\\
};
\addlegendentry{$G_{\text{es}}$ \, \,};
\addplot [
color=green,
line width=3.0pt,
mark size=5.0pt,
only marks,
mark=+,
mark options={solid}
]
table[row sep=crcr]{
0 0\\
0 0\\
0 0\\
0 0\\
0 0\\
0 0\\
0 0\\
0 0\\
12.5613262162514 1.23830904860254\\
0 0\\
0 0\\
14.8678514134985 0.110268336386102\\
0 0\\
0 0\\
0 0\\
18.1720805338954 0.738587729951646\\
0 0\\
0 0\\
0 0\\
0 0\\
0 0\\
0 0\\
24.6213482923394 0.119137591006705\\
0 0\\
0 0\\
0 0\\
0 0\\
0 0\\
0 0\\
29.2954070058747 0.0557892326749095\\
};
\addlegendentry{$G_{\text{rp}}$ \, \,};
\addplot [
color=red,
line width=3.0pt,
mark size=3.7pt,
only marks,
mark=*,
mark options={solid}
]
table[row sep=crcr]{
0 0\\
0 0\\
0 0\\
0 0\\
0 0\\
0 0\\
0 0\\
0 0\\
12.5613262162514 4.17615661806709\\
0 0\\
0 0\\
14.8678514134985 0.185345201667898\\
0 0\\
0 0\\
0 0\\
18.1720805338954 1.14459439972653\\
0 0\\
0 0\\
0 0\\
0 0\\
0 0\\
0 0\\
24.6213482923394 0.148443690211904\\
0 0\\
0 0\\
0 0\\
0 0\\
0 0\\
0 0\\
29.2954070058747 0.120749528280508\\
};
\addlegendentry{$G$};
\tikzstyle{every node}=[font=\LARGE]
\end{axis}
\begin{scope}[x={(image.south east)},y={(image.north west)},shift={(-18pt,0)}]
\node[scale=0.8] at (0.18,0.77) {$\times 10^{3}$};
\node at (0.62,0.8) {(a)};
\node at (1.2,0.8) {\textcolor{white}{(a)}};
\node[scale = 0.9] at (0.88,0.63) {\small{Forward}};
\node[anchor=south west,inner sep=0] at (0.32,0.47) {\raisebox{0pt}{\includegraphics[scale=0.05]{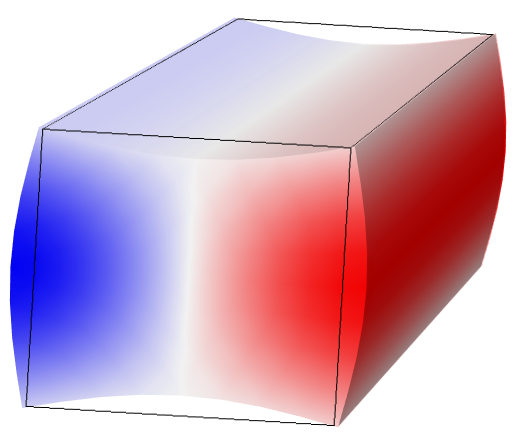}}};
\node[anchor=south west,inner sep=0] at (0.5,0.06) {\raisebox{0pt}{\includegraphics[scale=0.05]{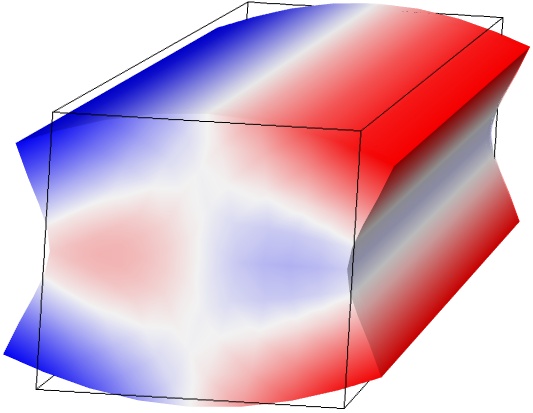}}};
\node[anchor=south west,inner sep=0] at (0.91,0.22) {\raisebox{0pt}{\includegraphics[scale=0.05]{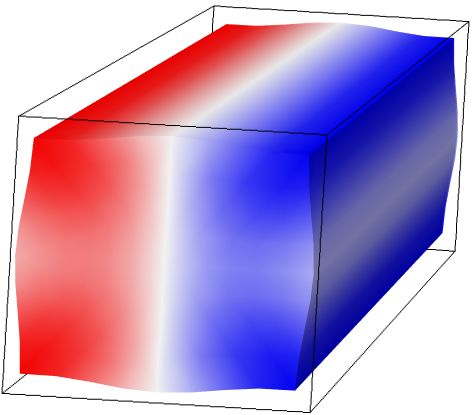}}};
\end{scope}
\end{tikzpicture}%
}
\subfigure{\begin{tikzpicture}[scale = 0.44*\columnwidth/(3.5in)]
\pgfplotsset{try min ticks=3}
\pgfplotsset{max space between ticks=50pt}
\begin{axis}[width=3.5in,
height=2.33in,
xmin=0,
xmax=0.5,
xlabel={$g$ ($\mu$m)},
ymode=log,
ymin=500,
ymax=100000,
yminorticks=false,
ylabel={Gain ($\text{W}^{-1}\text{m}^{-1}$)},
legend style={fill=none,draw=none,legend cell align=left,at={(0.5,-0.41)},anchor=north, legend columns=-1},
ytick = {1000,10000,100000}
]
\addplot [
color=blue,
loosely dotted,
line width=3.0pt
]
table[row sep=crcr]{
0.005 1076.2344164866\\
0.01 982.117640693436\\
0.015 934.769142426553\\
0.02 907.976546423901\\
0.025 891.974997985951\\
0.03 882.343818531904\\
0.035 876.734787779631\\
0.04 873.827385749191\\
0.045 872.71100468035\\
0.05 872.901333205304\\
0.055 874.097819646677\\
0.06 875.963505856904\\
0.065 878.350011683455\\
0.07 881.102517574257\\
0.075 884.15442282833\\
0.08 887.390653793902\\
0.085 890.772284796809\\
0.09 894.240016955614\\
0.095 897.78934211219\\
0.1 901.353749486579\\
0.105 904.862706543344\\
0.11 908.406637632437\\
0.115 911.949752082373\\
0.12 915.408250795786\\
0.125 918.869670873582\\
0.13 922.241680832767\\
0.135 925.600398729758\\
0.14 928.925257509273\\
0.145 932.098890766442\\
0.15 935.262132909912\\
0.155 938.395606105289\\
0.16 941.404831635356\\
0.165 944.414459656748\\
0.17 947.29814490123\\
0.175 950.12945202517\\
0.18 952.991431910042\\
0.185 955.67407254766\\
0.19 958.361096444261\\
0.195 960.896582980651\\
0.2 963.464396714346\\
0.205 965.971910960066\\
0.21 968.436793377405\\
0.215 970.832186149491\\
0.22 973.002745736911\\
0.225 975.382919657964\\
0.23 977.666787616537\\
0.235 979.801934150223\\
0.24 981.908526141663\\
0.245 984.089388284278\\
0.25 986.137412844451\\
0.255 988.057429345006\\
0.26 990.174907587015\\
0.265 991.946502242441\\
0.27 993.980867385398\\
0.275 995.768484556894\\
0.28 997.469928141439\\
0.285 999.212479302018\\
0.29 1000.96527661005\\
0.295 1002.67637816522\\
0.3 1004.27182701326\\
0.305 1006.02692494562\\
0.31 1007.54083053122\\
0.315 1009.06417296128\\
0.32 1010.72277199712\\
0.325 1012.15818880596\\
0.33 1013.81860457307\\
0.335 1015.1584961803\\
0.34 1016.51935302557\\
0.345 1017.93476678048\\
0.35 1019.06762992919\\
0.355 1020.34589293132\\
0.36 1021.98215768978\\
0.365 1023.32505172246\\
0.37 1024.18717545938\\
0.375 1025.3720456963\\
0.38 1026.80378949211\\
0.385 1027.68975155683\\
0.39 1028.93614708016\\
0.395 1030.38965087041\\
0.4 1031.50886247915\\
0.405 1032.15078780672\\
0.41 1033.6262661544\\
0.415 1034.20417501512\\
0.42 1035.19875142968\\
0.425 1036.54969430147\\
0.43 1037.5061842156\\
0.435 1037.96775271869\\
0.44 1039.26294796136\\
0.445 1040.09804876917\\
0.45 1040.67554409246\\
0.455 1042.1875966031\\
0.46 1042.47904611473\\
0.465 1043.62184229726\\
0.47 1044.40133534993\\
0.475 1044.8687648895\\
0.48 1046.4454357793\\
0.485 1046.73138128942\\
0.49 1047.09936260365\\
0.495 1048.18647359068\\
0.5 1048.17167808706\\
};
\addlegendentry{$G_{\text{es}}$ \, \,};
\addplot [
color=green,
densely dotted,
line width=3.0pt
]
table[row sep=crcr]{
0.005 26090.0454508476\\
0.01 11455.7814276226\\
0.015 6387.97391606046\\
0.02 4121.75129023859\\
0.025 2938.3012974522\\
0.03 2251.18112973505\\
0.035 1822.39419140087\\
0.04 1539.67035109967\\
0.045 1343.82522628619\\
0.05 1204.252333752\\
0.055 1102.62930325962\\
0.06 1027.00836965762\\
0.065 969.976058985987\\
0.07 926.411112906021\\
0.075 893.08958535365\\
0.08 867.628140329564\\
0.085 847.817416003462\\
0.09 832.963423097506\\
0.095 821.704219058959\\
0.1 813.708681669923\\
0.105 807.865441078431\\
0.11 804.218433842311\\
0.115 802.146216541673\\
0.12 801.21891152345\\
0.125 801.411510370563\\
0.13 802.373445157166\\
0.135 804.16661478488\\
0.14 806.685637049868\\
0.145 809.669302210752\\
0.15 812.972973603528\\
0.155 816.660797271721\\
0.16 820.505958213977\\
0.165 824.715651511854\\
0.17 828.987650570453\\
0.175 833.249947837495\\
0.18 837.834676949823\\
0.185 842.360372023528\\
0.19 846.979372787346\\
0.195 851.593359654191\\
0.2 856.200286642063\\
0.205 860.79135711156\\
0.21 865.427136570244\\
0.215 870.01543293362\\
0.22 874.2414061001\\
0.225 878.874091852961\\
0.23 883.320793548937\\
0.235 887.58357953601\\
0.24 891.900892763863\\
0.245 896.131729943424\\
0.25 900.344113056721\\
0.255 904.333976062531\\
0.26 908.390169912362\\
0.265 912.187110146918\\
0.27 916.065788685757\\
0.275 919.854129113262\\
0.28 923.431361500897\\
0.285 927.047092625884\\
0.29 930.562211835425\\
0.295 934.054587028013\\
0.3 937.356195961594\\
0.305 940.794578686789\\
0.31 943.931035321022\\
0.315 946.920664549921\\
0.32 950.057745547252\\
0.325 953.054187254259\\
0.33 956.164919539948\\
0.335 958.937622795879\\
0.34 961.717950432709\\
0.345 964.490238327805\\
0.35 966.891868330419\\
0.355 969.405921827474\\
0.36 972.164706884562\\
0.365 974.729149692041\\
0.37 976.820776767927\\
0.375 979.086514808753\\
0.38 981.610214535379\\
0.385 983.574924531804\\
0.39 985.846605207483\\
0.395 988.33979419578\\
0.4 990.263698404543\\
0.405 991.87962315772\\
0.41 994.21902965466\\
0.415 995.737191794005\\
0.42 997.575161085835\\
0.425 999.690565368188\\
0.43 1001.4269345965\\
0.435 1002.75422698122\\
0.44 1004.74771840931\\
0.445 1006.3770469683\\
0.45 1007.66533667488\\
0.455 1009.85568881356\\
0.46 1010.74980817499\\
0.465 1012.50485113465\\
0.47 1013.8422535249\\
0.475 1014.94720097135\\
0.48 1017.0230784558\\
0.485 1017.9236704105\\
0.49 1018.91457196267\\
0.495 1020.46624910489\\
0.5 1020.99636180615\\
};
\addlegendentry{$G_{\text{rp}}$ \, \,};
\addplot [
color=red,
solid,
line width=3.0pt
]
table[row sep=crcr]{
0.005 37764.2051061537\\
0.01 19146.3787852107\\
0.015 12209.9839567277\\
0.02 8898.81222536693\\
0.025 7068.10970305939\\
0.03 5952.25924168121\\
0.035 5227.17675139421\\
0.04 4733.33059101484\\
0.045 4382.42733011906\\
0.05 4127.70819454025\\
0.055 3940.19942457778\\
0.06 3799.93982677745\\
0.065 3694.37968066349\\
0.07 3614.45930082044\\
0.075 3554.4655552856\\
0.08 3509.92631612802\\
0.085 3476.64868339428\\
0.09 3453.31957240821\\
0.095 3437.30296880606\\
0.1 3427.88392940197\\
0.105 3422.70745689667\\
0.11 3422.07804059524\\
0.115 3424.67136106835\\
0.12 3429.45219985481\\
0.125 3436.54774651715\\
0.13 3445.05952755288\\
0.135 3455.2662911616\\
0.14 3466.91175013056\\
0.145 3479.22824544261\\
0.15 3492.18790093254\\
0.155 3505.88581121573\\
0.16 3519.6687634603\\
0.165 3534.20618476207\\
0.17 3548.62714355497\\
0.175 3562.92464102833\\
0.18 3577.94587931768\\
0.185 3592.49477267486\\
0.19 3607.24177659575\\
0.195 3621.68108718616\\
0.2 3636.16515694428\\
0.205 3650.49600305991\\
0.21 3664.83250536545\\
0.215 3678.93247538191\\
0.22 3691.84631877957\\
0.225 3706.00081422585\\
0.23 3719.58210678302\\
0.235 3732.49259688479\\
0.24 3745.45585034304\\
0.245 3758.38375862087\\
0.25 3771.01119063929\\
0.255 3782.92985109722\\
0.26 3795.36781109708\\
0.265 3806.59603073131\\
0.27 3818.5034852606\\
0.275 3829.7404283937\\
0.28 3840.37519027225\\
0.285 3851.16686720468\\
0.29 3861.77147486035\\
0.295 3872.24585310276\\
0.3 3882.10262417997\\
0.305 3892.54982691064\\
0.31 3901.90674996557\\
0.315 3910.98224644452\\
0.32 3920.6223459957\\
0.325 3929.53577102878\\
0.33 3939.12321897441\\
0.335 3947.3915100296\\
0.34 3955.71540322771\\
0.345 3964.1294692646\\
0.35 3971.23348908031\\
0.355 3978.85146125212\\
0.36 3987.6713663267\\
0.365 3995.51735003477\\
0.37 4001.4552145044\\
0.375 4008.38265348175\\
0.38 4016.31946802633\\
0.385 4022.04548943415\\
0.39 4029.10468045528\\
0.395 4037.02089625229\\
0.4 4043.12436707247\\
0.405 4047.66015422447\\
0.41 4055.30765389277\\
0.415 4059.51823007023\\
0.42 4065.1996171486\\
0.425 4072.14688810669\\
0.43 4077.54699856858\\
0.435 4081.14012459948\\
0.44 4087.72989932955\\
0.445 4092.67235199386\\
0.45 4096.41575489401\\
0.455 4103.83184493302\\
0.46 4106.21253263391\\
0.465 4112.01791436366\\
0.47 4116.26030738377\\
0.475 4119.41459501847\\
0.48 4126.72725567523\\
0.485 4129.10911955997\\
0.49 4131.83561017115\\
0.495 4137.11970968624\\
0.5 4138.15761925327\\
};
\addlegendentry{$G$};
\tikzstyle{every node}=[font=\LARGE]
\end{axis}
\begin{scope}[x={(image.south east)},y={(image.north west)},,shift={(-18pt,0)}]
\node[scale = 0.9] at (0.88,0.63) {\small{Forward}};
\node at (0.62,0.8) {(b)};
\node[anchor=south west,inner sep=0] at (0.5,0.33) {\raisebox{0pt}{\includegraphics[scale=0.05]{forward_mode1.png}}};
\end{scope}
\end{tikzpicture}
}
\subfigure{\begin{tikzpicture}[scale = 0.44*\columnwidth/(3.5in)]
\vspace{-100mm}
\begin{axis}[%
width=3.5in,
height=2.33in,
xmin=11,
xmax=19,
xlabel={Frequency (GHz)},
ymin=0,
ymax=0.75,
ylabel={Gain ($\text{W}^{-1}\text{m}^{-1}$)},
legend style={fill=none,draw=none,legend cell align=left},
xtick = {10,12,14,16,18,20}
]
\addplot [
color=black,
solid,
line width=2.0pt,
forget plot
]
table[row sep=crcr]{
12.2693029279845 0\\
12.2693029279845 0.716625018360622\\
};
\addplot [
color=black,
solid,
line width=2.0pt,
forget plot
]
table[row sep=crcr]{
16.4950972066937 0\\
16.4950972066937 0.39338175207008\\
};
\addplot [
color=black,
solid,
line width=2.0pt,
forget plot
]
table[row sep=crcr]{
14.8758777833039 0\\
14.8758777833039 0.214054857031902\\
};
\addplot [
color=black,
solid,
line width=2.0pt,
forget plot
]
table[row sep=crcr]{
22.0908681725107 0\\
22.0908681725107 0.170127706763883\\
};
\addplot [
color=black,
solid,
line width=2.0pt,
forget plot
]
table[row sep=crcr]{
25.0153962840304 0\\
25.0153962840304 0.169926435192134\\
};
\addplot [
color=black,
solid,
line width=2.0pt,
forget plot
]
table[row sep=crcr]{
0 0\\
0 0\\
};
\addplot [
color=black,
solid,
line width=2.0pt,
forget plot
]
table[row sep=crcr]{
0 0\\
0 0\\
};
\addplot [
color=black,
solid,
line width=2.0pt,
forget plot
]
table[row sep=crcr]{
0 0\\
0 0\\
};
\addplot [
color=black,
solid,
line width=2.0pt,
forget plot
]
table[row sep=crcr]{
0 0\\
0 0\\
};
\addplot [
color=black,
solid,
line width=2.0pt,
forget plot
]
table[row sep=crcr]{
0 0\\
0 0\\
};
\addplot [
color=blue,
line width=3.0pt,
mark size=5.0pt,
only marks,
mark=x,
mark options={solid}
]
table[row sep=crcr]{
0 0\\
12.2693029279845 0.0386266458723228\\
0 0\\
14.8758777833039 0.124956132376733\\
0 0\\
0 0\\
16.4950972066937 0.272554066980227\\
0 0\\
0 0\\
0 0\\
0 0\\
22.0908681725107 0.000496671036377048\\
0 0\\
0 0\\
0 0\\
0 0\\
25.0153962840304 0.142902460099527\\
0 0\\
0 0\\
0 0\\
0 0\\
0 0\\
0 0\\
0 0\\
0 0\\
0 0\\
0 0\\
0 0\\
0 0\\
0 0\\
};
\addplot [
color=green,
line width=3.0pt,
mark size=5.0pt,
only marks,
mark=+,
mark options={solid}
]
table[row sep=crcr]{
0 0\\
12.2693029279845 0.422500581244699\\
0 0\\
14.8758777833039 0.0119179320004351\\
0 0\\
0 0\\
16.4950972066937 0.0110532451410369\\
0 0\\
0 0\\
0 0\\
0 0\\
22.0908681725107 0.152239872973111\\
0 0\\
0 0\\
0 0\\
0 0\\
25.0153962840304 0.0011694296892263\\
0 0\\
0 0\\
0 0\\
0 0\\
0 0\\
0 0\\
0 0\\
0 0\\
0 0\\
0 0\\
0 0\\
0 0\\
0 0\\
};
\addplot [
color=red,
line width=3.0pt,
mark size=3.7pt,
only marks,
mark=*,
mark options={solid}
]
table[row sep=crcr]{
0 0\\
12.2693029279845 0.716625018360622\\
0 0\\
14.8758777833039 0.214054857031902\\
0 0\\
0 0\\
16.4950972066937 0.39338175207008\\
0 0\\
0 0\\
0 0\\
0 0\\
22.0908681725107 0.170127706763883\\
0 0\\
0 0\\
0 0\\
0 0\\
25.0153962840304 0.169926435192134\\
0 0\\
0 0\\
0 0\\
0 0\\
0 0\\
0 0\\
0 0\\
0 0\\
0 0\\
0 0\\
0 0\\
0 0\\
0 0\\
};
\tikzstyle{every node}=[font=\LARGE]
\end{axis}
\begin{scope}[x={(image.south east)},y={(image.north west)},shift={(-18pt,0)}]
\node[scale=0.8] at (0.18,0.77) {$\times 10^{3}$};
\node at (0.62,0.8) {(c)};
\node at (1.15,0.8) {\textcolor{white}{(a)}};
\node[scale = 0.9] at (0.87,0.63) {\small{Backward}};
\node[anchor=south west,inner sep=0] at (0.29,0.48) {\raisebox{0pt}{\includegraphics[scale=0.05]{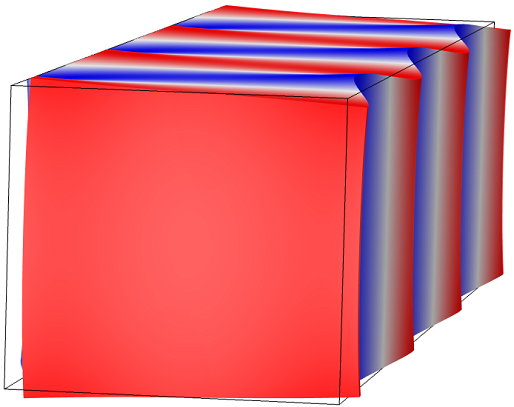}}};
\node[anchor=south west,inner sep=0] at (0.33,0.1) {\raisebox{0pt}{\includegraphics[scale=0.05]{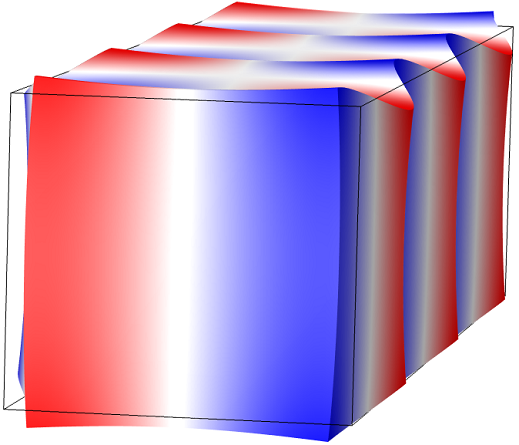}}};
\node[anchor=south west,inner sep=0] at (0.85,0.28) {\raisebox{0pt}{\includegraphics[scale=0.05]{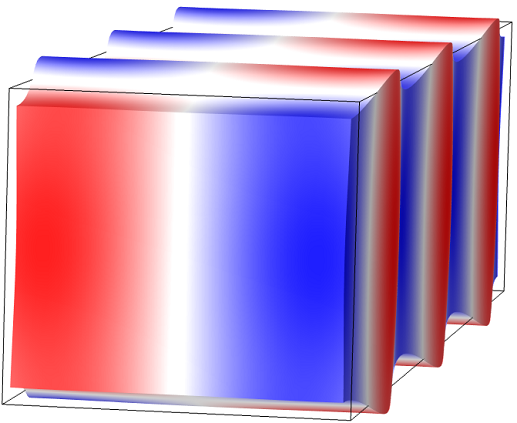}}};
\end{scope}
\end{tikzpicture}
}
\subfigure{\begin{tikzpicture}[scale = 0.44*\columnwidth/(3.5in)]
\pgfplotsset{try min ticks=3}
\pgfplotsset{max space between ticks=50pt}
\begin{axis}[width=3.5in,
height=2.33in,
xmin=0,
xmax=0.5,
xlabel={$g$ ($\mu$m)},
ymode=log,
ymin=0.1,
ymax=30000,
yminorticks=true,
ylabel={Gain ($\text{W}^{-1}\text{m}^{-1}$)},
legend style={fill=none,draw=none,legend cell align=left},
ytick = {1,100,10000},
extra y ticks = {0.1,10,1000},
extra y tick labels={}
]
\addplot [
color=blue,
loosely dotted,
line width=3.0pt
]
table[row sep=crcr]{
0.005 231.566078177385\\
0.01 155.556014565747\\
0.015 116.669240385805\\
0.02 92.9679320224054\\
0.025 76.9053323299598\\
0.03 65.243881955012\\
0.035 56.3468503399336\\
0.04 49.3041433683248\\
0.045 43.5758935333504\\
0.05 38.8087502810878\\
0.055 34.7736233709742\\
0.06 31.3113009990485\\
0.065 28.3060409830036\\
0.07 25.6763215480028\\
0.075 23.3552046338546\\
0.08 21.293200276886\\
0.085 19.4516065975762\\
0.09 17.8001895091189\\
0.095 16.3134549692544\\
0.1 14.9692531262772\\
0.105 13.7530058498916\\
0.11 12.64750855942\\
0.115 11.6410265518741\\
0.12 10.7242418677725\\
0.125 9.88631285204308\\
0.13 9.1192567193778\\
0.135 8.41784466428681\\
0.14 7.77414209942788\\
0.145 7.18298140226754\\
0.15 6.64003425631152\\
0.155 6.14064228452636\\
0.16 5.68137018942941\\
0.165 5.25849036177031\\
0.17 4.86930203011279\\
0.175 4.50962745372511\\
0.18 4.17848283788549\\
0.185 3.87155929749139\\
0.19 3.59003579153535\\
0.195 3.32828441314989\\
0.2 3.08770787712761\\
0.205 2.86414115321651\\
0.21 2.65754660432459\\
0.215 2.46645264827104\\
0.22 2.29034934558804\\
0.225 2.12659500363728\\
0.23 1.97494916271275\\
0.235 1.83464270877961\\
0.24 1.70439588153619\\
0.245 1.58349231061869\\
0.25 1.47172793506891\\
0.255 1.36837694088944\\
0.26 1.27234938961329\\
0.265 1.18292733919579\\
0.27 1.10008103389073\\
0.275 1.02315289436201\\
0.28 0.95210778452289\\
0.285 0.885867548972481\\
0.29 0.82412490558788\\
0.295 0.766914937337702\\
0.3 0.713671974211493\\
0.305 0.664323258683818\\
0.31 0.618825730927889\\
0.315 0.576012132275737\\
0.32 0.536611764298678\\
0.325 0.499732587335969\\
0.33 0.465366917308543\\
0.335 0.433537548674106\\
0.34 0.403914543231447\\
0.345 0.376179137543114\\
0.35 0.350968052720649\\
0.355 0.327161010121978\\
0.36 0.304885951825078\\
0.365 0.284100782941135\\
0.37 0.264962814096608\\
0.375 0.246992578486478\\
0.38 0.230385805732191\\
0.385 0.214669213392995\\
0.39 0.200186002725318\\
0.395 0.187060147680231\\
0.4 0.174419086099597\\
0.405 0.162746581130915\\
0.41 0.151893771340866\\
0.415 0.14170757049003\\
0.42 0.13220857089637\\
0.425 0.123385239214889\\
0.43 0.115137573596893\\
0.435 0.107501954434036\\
0.44 0.100368043441469\\
0.445 0.0936572420586102\\
0.45 0.0875122798391527\\
0.455 0.0816620114533059\\
0.46 0.0765381545311223\\
0.465 0.0715458095948337\\
0.47 0.0665976755733651\\
0.475 0.0625071714630435\\
0.48 0.058339506489705\\
0.485 0.054505670196911\\
0.49 0.0509812946080245\\
0.495 0.0476558782272866\\
0.5 0.0444928719046404\\
};
\addplot [
color=green,
densely dotted,
line width=3.0pt
]
table[row sep=crcr]{
0.005 13734.6326375009\\
0.01 6573.16345179432\\
0.015 3769.09198758319\\
0.02 2403.03963837617\\
0.025 1645.23345673579\\
0.03 1185.91535000419\\
0.035 886.053767178299\\
0.04 681.370665709186\\
0.045 536.640820352542\\
0.05 430.716679583919\\
0.055 351.009214220677\\
0.06 289.780667241551\\
0.065 241.42098426535\\
0.07 203.338878461777\\
0.075 173.137090518578\\
0.08 148.126757246689\\
0.085 127.568507679488\\
0.09 110.295198341611\\
0.095 96.085364295019\\
0.1 84.0103639420629\\
0.105 73.7295435652214\\
0.11 64.8543156912638\\
0.115 57.2559263736805\\
0.12 50.876499326987\\
0.125 45.2308077922112\\
0.13 40.3092697996631\\
0.135 35.955816762071\\
0.14 32.193025728382\\
0.145 28.90320198051\\
0.15 25.9279347855983\\
0.155 23.356327433294\\
0.16 20.9886008424965\\
0.165 18.9663438602369\\
0.17 17.1583926259115\\
0.175 15.4599825769253\\
0.18 14.0060478669095\\
0.185 12.7186913358404\\
0.19 11.512784325893\\
0.195 10.4910219261\\
0.2 9.48124576208069\\
0.205 8.63810232118966\\
0.21 7.86975860678887\\
0.215 7.16497883844846\\
0.22 6.53997341864043\\
0.225 5.97982420519086\\
0.23 5.44114787640619\\
0.235 4.96248300980285\\
0.24 4.53058324241091\\
0.245 4.14737290986694\\
0.25 3.77969204660023\\
0.255 3.44791527358179\\
0.26 3.15649455446347\\
0.265 2.88714541640093\\
0.27 2.63605291908206\\
0.275 2.41787815720595\\
0.28 2.20485969612521\\
0.285 2.01735153873888\\
0.29 1.84671757266538\\
0.295 1.69342207300684\\
0.3 1.54398478427807\\
0.305 1.41702206805275\\
0.31 1.28366758040007\\
0.315 1.17884770065209\\
0.32 1.07707681518793\\
0.325 0.987205170280872\\
0.33 0.904485612499634\\
0.335 0.828137401642403\\
0.34 0.756112362421344\\
0.345 0.69102242239186\\
0.35 0.626185121848518\\
0.355 0.574429028561644\\
0.36 0.521377613273035\\
0.365 0.475515119288606\\
0.37 0.436587687074483\\
0.375 0.396113377604886\\
0.38 0.362153836705983\\
0.385 0.331835021917783\\
0.39 0.298279904420357\\
0.395 0.264558902593773\\
0.4 0.240126077635499\\
0.405 0.217400237683198\\
0.41 0.197179767364918\\
0.415 0.179355488321168\\
0.42 0.162312844404561\\
0.425 0.14732641620124\\
0.43 0.1334109810922\\
0.435 0.120814621740771\\
0.44 0.107645494279423\\
0.445 0.0979043966086994\\
0.45 0.0874346174688838\\
0.455 0.0780403755331599\\
0.46 0.0636402056548943\\
0.465 0.0564998059958836\\
0.47 0.0478217743763976\\
0.475 0.0443217310118969\\
0.48 0.0396662360294981\\
0.485 0.0336413202501604\\
0.49 0.0298717488476877\\
0.495 0.026329415147942\\
0.5 0.0225611719132711\\
};
\addplot [
color=red,
solid,
line width=3.0pt
]
table[row sep=crcr]{
0.005 17532.9759809828\\
0.01 8751.0894717254\\
0.015 5212.01474692464\\
0.02 3441.32365668443\\
0.025 2433.55210283105\\
0.03 1807.48187613005\\
0.035 1389.28466964956\\
0.04 1097.25029646911\\
0.045 886.057216457623\\
0.05 728.102892609707\\
0.055 606.743125562602\\
0.06 511.600862476069\\
0.065 435.059082330725\\
0.07 373.528098072131\\
0.075 323.67173026317\\
0.08 281.742576944943\\
0.085 246.64766921286\\
0.09 216.713115928729\\
0.095 191.581753375891\\
0.1 169.904151697181\\
0.105 151.169372038394\\
0.11 134.781681541728\\
0.115 120.530963390292\\
0.12 108.317414199937\\
0.125 97.4097165156088\\
0.13 87.7738302457708\\
0.135 79.168516234543\\
0.14 71.6072156547637\\
0.145 64.9036199302559\\
0.15 58.8101012092306\\
0.155 53.4488257501305\\
0.16 48.5097522711593\\
0.165 44.19825019295\\
0.17 40.3087664021581\\
0.175 36.6691623435736\\
0.18 33.4847307869486\\
0.185 30.6246608729103\\
0.19 27.9607060696653\\
0.195 25.6374457452059\\
0.2 23.3902876095762\\
0.205 21.4502572146087\\
0.21 19.6737253331778\\
0.215 18.0390663576917\\
0.22 16.5708198644272\\
0.225 15.2385056573755\\
0.23 13.972312545859\\
0.235 12.8318198355622\\
0.24 11.7926431725639\\
0.245 10.8562269099833\\
0.25 9.96848617621817\\
0.255 9.16050572419188\\
0.26 8.43691775561731\\
0.265 7.76616974853101\\
0.27 7.14193595579943\\
0.275 6.58673226736068\\
0.28 6.05473495265932\\
0.285 5.57687296436666\\
0.29 5.13816973277005\\
0.295 4.73955677421667\\
0.3 4.3570839843283\\
0.305 4.02182023549029\\
0.31 3.68503815363574\\
0.315 3.40292606664655\\
0.32 3.13417809287786\\
0.325 2.89169908108201\\
0.33 2.66741590974595\\
0.335 2.46005495527886\\
0.34 2.26529570463169\\
0.345 2.08690390646228\\
0.35 1.91474791187909\\
0.355 1.76860971833943\\
0.36 1.6236611068125\\
0.365 1.49471920841259\\
0.37 1.38178429088514\\
0.375 1.26868429493509\\
0.38 1.17024234614835\\
0.385 1.08030106232022\\
0.39 0.987184483666769\\
0.395 0.89653894144234\\
0.4 0.823849798499301\\
0.405 0.756344348949486\\
0.41 0.695197091774176\\
0.415 0.639911176284365\\
0.42 0.587500248644582\\
0.425 0.540362578690368\\
0.43 0.4964244644031\\
0.435 0.456244704858756\\
0.44 0.415899733954463\\
0.445 0.383076189255855\\
0.45 0.349893777378145\\
0.455 0.319363704037393\\
0.46 0.279762082658326\\
0.465 0.255204167987739\\
0.47 0.227287851466963\\
0.475 0.212098575978747\\
0.48 0.194216108577442\\
0.485 0.173789099401578\\
0.49 0.158901757178896\\
0.495 0.144830233973298\\
0.5 0.130420007298497\\
};
\tikzstyle{every node}=[font=\LARGE]
\end{axis}
\begin{scope}[x={(image.south east)},y={(image.north west)},,shift={(-18pt,0)}]
\node[scale = 0.9] at (0.87,0.63) {\small{Backward}};
\node at (0.62,0.8) {(d)};
\node[anchor=south west,inner sep=0] at (0.5,0.33) {\raisebox{0pt}{\includegraphics[scale=0.05]{backward_mode1.png}}};
\end{scope}
\end{tikzpicture}
}
\vspace{-6mm}
\caption{(a-c) Brillouin spectrum of a vertical slot waveguide and (b-d) the gain of the most promising mode increases rapidly in narrow slots. The color of the modes indicates the sign of $u_{x}$ (red: $+$, blue: $-$).}
\vspace{-6mm}
\end{figure}

In the forward case (fig.4a), the mechanical modes are identical to those of a stand-alone wire. The maximum gain among all modes is $4.2\times10^{3} \, \text{W}^{-1}\text{m}^{-1}$. This is smaller than $\tilde{G} = 1.7 \times10^{4} \, \text{W}^{-1}\text{m}^{-1}$ \cite{Qiu2012}, despite the increase in radiation pressure close to the slot. The cause is a decrease in the pressure on the far side from the slot (fig.3a). If these two effects would perfectly balance, we would expect $G = \tilde{G}/4$. This explains why $G \approx \tilde{G}/4$ in slots as narrow as $50 \, \text{nm}$. Hence, smaller gaps are necessary to boost $G$  substantially. Indeed, for the most promising mode we numerically find that $G \propto 1/g$ as $g$ falls below $50 \, \text{nm}$ (fig.4b). Eventually $G$ approaches a maximum of $\, \approx 1.1 \times10^{5} \, \text{W}^{-1}\text{m}^{-1}$ as $g \rightarrow 0$. In wide slots, the optical mode evolves into the symmetric supermode of two weakly coupled silicon wires. Therefore $G \rightarrow \tilde{G}/4$ as $g \rightarrow \infty$.

In the backward case, the mechanical modes are different from those of a stand-alone wire since the phonon wavevector $K \approx 2\beta$ depends on the effective index $n_{p}$ of the optical mode. From the point of view of a single beam, horizontal symmetry is broken by the slot waveguide. So modes that were previously forbidden by symmetry can have non-zero gain in the slot waveguide. Such a previously forbidden phonon has the largest backward SBS gain in the slot waveguide (fig.4c). For $g = 50 \, \text{nm}$, this phonon has a gain of $7.2\times10^{2} \, \text{W}^{-1}\text{m}^{-1}$. The optical forces are symmetric again in wide slots. Then this mode is forbidden, which means that $G \rightarrow 0$ as $g \rightarrow \infty$ (fig.4d). Going from wide to narrow slots, $G$ first increases exponentially, then its growth accelerates like $G \propto 1/g^{1.6}$ and ultimately converges to a maximum of $\, \approx4.5 \times10^{4} \, \text{W}^{-1}\text{m}^{-1}$ as $g \rightarrow 0$.

In general, gradient forces dominate the SBS gain in narrow slots (fig.4b-d). The slot enhances these forces despite the reduced dispersion in such waveguides. As $g \rightarrow 0$, the group and effective indices $n_{g}$ and $n_{p}$ approach those of a single wire of width $a+\bar{a}$. Thus the waveguide dispersion decreases (fig.5a), contrary to the prediction that very dispersive waveguides are optimal for large gradient forces \cite{Rakich2011}. Writing the  gradient force density as $\mathbf{p}\delta\left(\mathbf{r}-\mathbf{r}_{\partial \text{wg}}\right)$, it was shown that $c \int \mathbf{p} \cdot \mathbf{r}dl = n_{\text{g}} - n_{\text{p}}$ from the scale-invariance of Maxwell's equations \cite{Rakich2011}. For a stand-alone wire the integral becomes $\int \mathbf{p} \cdot \mathbf{r}dl = A_{\text{wg}}\left(\bar{p}_{x} + \bar{p}_{y}\right)$ with $A_{\text{wg}} = ab$ and $\bar{p}$ the magnitude of the spatially averaged radiation pressure. However, this no longer holds for a slot waveguide. Then the integral yields $\int \mathbf{p} \cdot \mathbf{r}dl = A_{\text{g}}\left(\bar{p}_{x,L} - \bar{p}_{x,R}\right)+2A_{\text{wg}}\left(\bar{p}_{x,L} + \bar{p}_{y}\right)$, with $\bar{p}_{x,L/R}$ the pressure on the left/right boundary, $A_{g} = gb$ and $a = \bar{a}$. Since $A_{\text{g}} \rightarrow 0$ as $g \rightarrow 0$, $\bar{p}_{x,R}$ and thus $\bar{p}_{x}+\bar{p}_{y}$ can increase drastically in narrow slots (fig.5a).

\begin{figure}
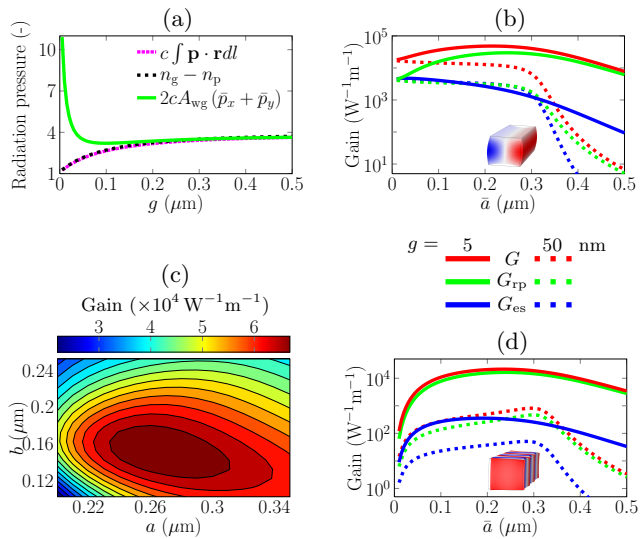

\centering
\label{fig:e}
\subfigure{\definecolor{mycolor1}{rgb}{1,0,1}%
\begin{tikzpicture}[scale = 0.44*\columnwidth/(3.5in)]
\pgfplotsset{try min ticks=3}
\pgfplotsset{max space between ticks=50pt}
\begin{axis}[%
width=3.5in,
height=2.33in,
xmin=0,
xmax=0.5,
xlabel={$g$ ($\mu$m)},
ymin=1,
ymax=11,
legend style={fill=none,draw=none,legend cell align=left,row sep=1.2mm},
ylabel={Radiation pressure (-)},
ytick = {1,4,7,10}
]
\addplot [
color=mycolor1,
densely dotted,
line width=3.0pt
]
table[row sep=crcr]{
0.005 1.2611393080655\\
0.01 1.3723810917399\\
0.015 1.49000326953065\\
0.02 1.60097570331833\\
0.025 1.70749481901461\\
0.03 1.80218366643589\\
0.035 1.89150121648344\\
0.04 1.98017719780016\\
0.045 2.0558998211794\\
0.05 2.1279411948821\\
0.055 2.1971888861859\\
0.06 2.26083704422338\\
0.065 2.32085196160949\\
0.07 2.37722486402238\\
0.075 2.43055362203945\\
0.08 2.4819526839472\\
0.085 2.52710808932888\\
0.09 2.57252282273488\\
0.095 2.6156990385429\\
0.1 2.6563911824433\\
0.105 2.69514745764425\\
0.11 2.7322385109661\\
0.115 2.76852699836172\\
0.12 2.80186203172702\\
0.125 2.83378919782402\\
0.13 2.86541134117403\\
0.135 2.89442087443795\\
0.14 2.92205184503985\\
0.145 2.94892706634178\\
0.15 2.9741314806464\\
0.155 2.99909816953602\\
0.16 3.02240099232793\\
0.165 3.0453186458869\\
0.17 3.0675798364056\\
0.175 3.08798475089994\\
0.18 3.10878744133152\\
0.185 3.12802040666857\\
0.19 3.14637230564385\\
0.195 3.16423646630486\\
0.2 3.18143369360203\\
0.205 3.19819766104923\\
0.21 3.21423105407048\\
0.215 3.22969851740359\\
0.22 3.24423969642287\\
0.225 3.259036696748\\
0.23 3.27342774566723\\
0.235 3.28601173251576\\
0.24 3.29927126387151\\
0.245 3.31124789269544\\
0.25 3.32412284000368\\
0.255 3.33565653941592\\
0.26 3.34647094380051\\
0.265 3.35742794470705\\
0.27 3.36871596621266\\
0.275 3.37928282110044\\
0.28 3.38865450233177\\
0.285 3.39841278334572\\
0.29 3.40729274392207\\
0.295 3.41677035241431\\
0.3 3.42509483603479\\
0.305 3.43418908872072\\
0.31 3.44195316760015\\
0.315 3.45025108539498\\
0.32 3.45807041793464\\
0.325 3.46536464622117\\
0.33 3.47295887381765\\
0.335 3.47989235957355\\
0.34 3.48690452770672\\
0.345 3.4937400697503\\
0.35 3.50013611528451\\
0.355 3.50650301367256\\
0.36 3.5129884825812\\
0.365 3.51918709692801\\
0.37 3.52433820974911\\
0.375 3.5299433309861\\
0.38 3.53597817856255\\
0.385 3.54043670519473\\
0.39 3.54548291740206\\
0.395 3.5516785737788\\
0.4 3.5568771833303\\
0.405 3.56059627395408\\
0.41 3.56611496203896\\
0.415 3.57018273743123\\
0.42 3.57421171670801\\
0.425 3.57928263771752\\
0.43 3.58269367341553\\
0.435 3.58643062258004\\
0.44 3.59078445905134\\
0.445 3.59480563996018\\
0.45 3.59775560289338\\
0.455 3.60283156033133\\
0.46 3.60515282560526\\
0.465 3.60984273912539\\
0.47 3.61232291946958\\
0.475 3.6155275166203\\
0.48 3.61983169692831\\
0.485 3.62255256046839\\
0.49 3.62487831586378\\
0.495 3.62823406443894\\
0.5 3.62998353037636\\
};
\addlegendentry{$c \int \mathbf{p} \cdot \mathbf{r}dl$};
\addplot [
color=black,
loosely dotted,
line width=3.0pt
]
table[row sep=crcr]{
0.005 1.28420590979843\\
0.01 1.39785671471113\\
0.015 1.51862901862896\\
0.02 1.63404924193311\\
0.025 1.74160871986362\\
0.03 1.84126113174528\\
0.035 1.93357010689872\\
0.04 2.01918226100166\\
0.045 2.09897381885007\\
0.05 2.17336851847816\\
0.055 2.24301759998675\\
0.06 2.30851444385656\\
0.065 2.36981629186581\\
0.07 2.42766307292137\\
0.075 2.48234409369363\\
0.08 2.53397027768609\\
0.085 2.5828534082891\\
0.09 2.62916257997061\\
0.095 2.67315824394183\\
0.1 2.71492089017236\\
0.105 2.75447879516488\\
0.11 2.79250650333962\\
0.115 2.82865166491253\\
0.12 2.86301601530188\\
0.125 2.89589286449428\\
0.13 2.92725741217723\\
0.135 2.95725330899597\\
0.14 2.98598138962993\\
0.145 3.01349540464395\\
0.15 3.03982187975025\\
0.155 3.06502561379658\\
0.16 3.08928484189448\\
0.165 3.11254058140888\\
0.17 3.13484790317116\\
0.175 3.15627968238739\\
0.18 3.17690082320732\\
0.185 3.19669137305778\\
0.19 3.21577412539322\\
0.195 3.23415899983657\\
0.2 3.2518547257059\\
0.205 3.26889655446871\\
0.21 3.2853431592281\\
0.215 3.30119957505741\\
0.22 3.31642303471558\\
0.225 3.33128202022204\\
0.23 3.34547100586695\\
0.235 3.35920062872777\\
0.24 3.37253530012191\\
0.245 3.38533848337543\\
0.25 3.39780281472208\\
0.255 3.40982287625773\\
0.26 3.42146236051977\\
0.265 3.43269200057896\\
0.27 3.44360435052233\\
0.275 3.45413203750726\\
0.28 3.46437169306482\\
0.285 3.47426788465013\\
0.29 3.483873146091\\
0.295 3.49313079298721\\
0.3 3.50216470085685\\
0.305 3.51091164036323\\
0.31 3.51938777657734\\
0.315 3.52760949664184\\
0.32 3.53558545898285\\
0.325 3.54332589586173\\
0.33 3.55089134896394\\
0.335 3.55810903708515\\
0.34 3.56523205815626\\
0.345 3.5721154917854\\
0.35 3.57880417087824\\
0.355 3.58523283206882\\
0.36 3.5916139757133\\
0.365 3.59771764939436\\
0.37 3.60363544486837\\
0.375 3.60936266957155\\
0.38 3.61508073522154\\
0.385 3.62054355772685\\
0.39 3.62589109791258\\
0.395 3.63100409857406\\
0.4 3.63602432250964\\
0.405 3.64088942096692\\
0.41 3.64561033394013\\
0.415 3.65029371259951\\
0.42 3.65478349537005\\
0.425 3.65918215242709\\
0.43 3.6633927161047\\
0.435 3.66757845375005\\
0.44 3.67157833479323\\
0.445 3.67548346288651\\
0.45 3.67930390535711\\
0.455 3.68304344690227\\
0.46 3.68661185213961\\
0.465 3.69012366824275\\
0.47 3.69355196055632\\
0.475 3.69684285974616\\
0.48 3.7000760641495\\
0.485 3.70321820816057\\
0.49 3.70632084029292\\
0.495 3.70930082273371\\
0.5 3.71219283370781\\
};
\addlegendentry{$n_{\text{g}} - n_{\text{p}}$};
\addplot [
color=green,
solid,
line width=3.0pt
]
table[row sep=crcr]{
0.005 10.9131226986637\\
0.01 7.70070987017594\\
0.015 6.09908977913374\\
0.02 5.17135015701534\\
0.025 4.58791540613358\\
0.03 4.1933621064973\\
0.035 3.92144238566083\\
0.04 3.73283001100557\\
0.045 3.58875908233909\\
0.05 3.48208408237321\\
0.055 3.40524942121627\\
0.06 3.34702091691658\\
0.065 3.30375507387424\\
0.07 3.27115990352652\\
0.075 3.24783865869053\\
0.08 3.23265189905122\\
0.085 3.21827439447148\\
0.09 3.21168254351617\\
0.095 3.20811350542847\\
0.1 3.20788348301362\\
0.105 3.20910732983223\\
0.11 3.21320264250979\\
0.115 3.21933491571729\\
0.12 3.22494283471507\\
0.125 3.2315220007471\\
0.13 3.23912063773247\\
0.135 3.24688592097854\\
0.14 3.25552001215542\\
0.145 3.26471309940502\\
0.15 3.27338691581798\\
0.155 3.28307932637621\\
0.16 3.29170828483481\\
0.165 3.30166721121072\\
0.17 3.31129800885049\\
0.175 3.31990175534957\\
0.18 3.32944961430329\\
0.185 3.33853692684955\\
0.19 3.34741821706496\\
0.195 3.3565086237531\\
0.2 3.36537956498485\\
0.205 3.37378940424333\\
0.21 3.38237316027453\\
0.215 3.3908774794568\\
0.22 3.39817025066623\\
0.225 3.4067035607722\\
0.23 3.41501739914007\\
0.235 3.42198080874858\\
0.24 3.42996366986146\\
0.245 3.43677669444175\\
0.25 3.44468636391384\\
0.255 3.451385121327\\
0.26 3.45785935947169\\
0.265 3.4644130004115\\
0.27 3.47123787710354\\
0.275 3.47773701292836\\
0.28 3.483626273027\\
0.285 3.48969609651434\\
0.29 3.49540665785769\\
0.295 3.50141452816799\\
0.3 3.50690326761071\\
0.305 3.51260111534767\\
0.31 3.51777830664662\\
0.315 3.52242914003413\\
0.32 3.52734391123231\\
0.325 3.53245295587143\\
0.33 3.53749632118855\\
0.335 3.54197988392898\\
0.34 3.54669272321396\\
0.345 3.55128801014483\\
0.35 3.55527492503014\\
0.355 3.55935633753662\\
0.36 3.56351245281885\\
0.365 3.56789937094775\\
0.37 3.57123235517343\\
0.375 3.57487904117562\\
0.38 3.57896153530316\\
0.385 3.58203564178548\\
0.39 3.5855391368782\\
0.395 3.58983277171977\\
0.4 3.59282835270979\\
0.405 3.59522916446504\\
0.41 3.59911791725271\\
0.415 3.60163865260153\\
0.42 3.60456721831789\\
0.425 3.60802159514452\\
0.43 3.61048897928148\\
0.435 3.61278353983046\\
0.44 3.61598511941069\\
0.445 3.61868989385224\\
0.45 3.62066050144364\\
0.455 3.62445952406368\\
0.46 3.62550770428763\\
0.465 3.62846674079307\\
0.47 3.63042815361807\\
0.475 3.63233972318126\\
0.48 3.63560311118723\\
0.485 3.63714867531924\\
0.49 3.63878381113152\\
0.495 3.64125773958648\\
0.5 3.64194088771133\\
};
\addlegendentry{$2cA_{\text{wg}}\left(\bar{p}_{x} + \bar{p}_{y}\right)$};
\tikzstyle{every node}=[font=\LARGE]
\end{axis}
\begin{scope}[x={(image.south east)},y={(image.north west)},,shift={(-18pt,0)}]
\node at (0.62,0.8) {(a)};
\node at (1.2,0.8) {\textcolor{white}{(a)}};
\end{scope}
\end{tikzpicture}%
}
\subfigure{\begin{tikzpicture}[scale = 0.44*\columnwidth/(3.5in)]
\pgfplotsset{try min ticks=3}
\pgfplotsset{max space between ticks=50pt}
\begin{axis}[%
width=3.5in,
height=2.33in,
xmin=0,
xmax=0.5,
xlabel={$\bar{a}$ ($\mu$m)},
ymode=log,
ymin=5,
ymax=100000,
yminorticks=false,
legend style={fill=none,draw=none,legend cell align=left,at={(0.2,-0.31)},anchor=north},
ylabel={Gain ($\text{W}^{-1}\text{m}^{-1}$)},
ytick = {10,1000,100000},
extra y ticks = {100,10000},
extra y tick labels={}
]
\addplot [
color=blue,
loosely dotted,
line width=3.0pt
]
table[row sep=crcr]{
0.01 4293.49357322311\\
0.02 4219.89971174495\\
0.03 4136.67379307368\\
0.04 4056.8319625907\\
0.05 3983.76099528798\\
0.06 3917.85249590844\\
0.07 3858.14602352902\\
0.08 3803.60050881437\\
0.09 3753.48239522265\\
0.1 3706.52989351563\\
0.11 3662.02184014059\\
0.12 3619.2960006168\\
0.13 3577.65293590107\\
0.14 3536.26600467762\\
0.15 3494.55748198594\\
0.16 3451.78203352009\\
0.17 3407.24787349341\\
0.18 3360.07256730516\\
0.19 3309.0378644538\\
0.2 3252.71976571067\\
0.21 3189.5173468717\\
0.22 3116.99125751487\\
0.23 3032.02205568384\\
0.24 2930.48385722759\\
0.25 2806.23504099156\\
0.26 2651.14065944284\\
0.27 2454.60954478992\\
0.28 2203.76213482972\\
0.29 1886.70796322088\\
0.3 1502.75965148258\\
0.31 1079.7401143897\\
0.32 682.30629476856\\
0.33 379.662499842671\\
0.34 193.091039064374\\
0.35 94.7761854007163\\
0.36 46.9683293487696\\
0.37 24.1182356756832\\
0.38 12.9693525726232\\
0.39 7.31605862497803\\
0.4 4.3187287803054\\
0.41 2.65715359209233\\
0.42 1.6966853848532\\
0.43 1.11953707132551\\
0.44 0.760597995368689\\
0.45 0.530256568758589\\
0.46 0.378259288440914\\
0.47 0.275426275684322\\
0.48 0.204255529488676\\
0.49 0.153991334681893\\
0.5 0.117837505680701\\
};
\addplot [
color=green,
loosely dotted,
line width=3.0pt
]
table[row sep=crcr]{
0.01 4084.41232222955\\
0.02 3940.82091444485\\
0.03 3829.99188937506\\
0.04 3742.77117157398\\
0.05 3671.60809787254\\
0.06 3612.7397826524\\
0.07 3562.22623806628\\
0.08 3518.22077406451\\
0.09 3479.51760968755\\
0.1 3443.86595916803\\
0.11 3411.2530510881\\
0.12 3380.21899125635\\
0.13 3350.85562250171\\
0.14 3321.73379676186\\
0.15 3293.03137438432\\
0.16 3263.30356594372\\
0.17 3233.00278556452\\
0.18 3200.9504883103\\
0.19 3166.05452634849\\
0.2 3126.920054813\\
0.21 3083.74656087543\\
0.22 3033.8120551131\\
0.23 2974.12359373783\\
0.24 2902.48926922082\\
0.25 2813.29576307114\\
0.26 2699.45359518661\\
0.27 2551.99534460934\\
0.28 2358.82286960841\\
0.29 2105.01328187171\\
0.3 1782.24521647554\\
0.31 1401.89182562606\\
0.32 1010.05371495018\\
0.33 672.412154375252\\
0.34 427.849657546297\\
0.35 271.245679100512\\
0.36 176.030072914375\\
0.37 118.341555011857\\
0.38 82.4426606014234\\
0.39 59.3161578841191\\
0.4 43.8807289517041\\
0.41 33.2319513556526\\
0.42 25.6569510568293\\
0.43 20.1507844341184\\
0.44 16.0484429690721\\
0.45 12.9441700940232\\
0.46 10.553238442595\\
0.47 8.68631859338039\\
0.48 7.21226966791049\\
0.49 6.03529320847268\\
0.5 5.08459342415143\\
};
\addplot [
color=red,
loosely dotted,
line width=3.0pt
]
table[row sep=crcr]{
0.01 16753.2024408619\\
0.02 16316.6679141274\\
0.03 15927.4262180917\\
0.04 15592.8806776644\\
0.05 15304.3714151076\\
0.06 15055.0009794575\\
0.07 14834.8416277216\\
0.08 14638.0788853292\\
0.09 14460.8096570329\\
0.1 14295.9657045337\\
0.11 14142.1031322686\\
0.12 13994.9458081312\\
0.13 13853.3041382373\\
0.14 13712.6432792807\\
0.15 13572.1853595587\\
0.16 13427.5255787401\\
0.17 13278.2147644484\\
0.18 13120.1162627817\\
0.19 12948.6059062319\\
0.2 12758.0392022141\\
0.21 12545.6360744057\\
0.22 12301.0441705882\\
0.23 12012.012225543\\
0.24 11665.8790743368\\
0.25 11239.0571723622\\
0.26 10700.9703851378\\
0.27 10012.2625409522\\
0.28 9122.53435626798\\
0.29 7977.46851334385\\
0.3 6558.09893954682\\
0.31 4942.26506192847\\
0.32 3352.68047006955\\
0.33 2062.59863603073\\
0.34 1195.79349494761\\
0.35 686.693851251386\\
0.36 404.853712220597\\
0.37 249.309021176164\\
0.38 160.810115053688\\
0.39 108.295653164691\\
0.4 75.7319089141695\\
0.41 54.6829761739622\\
0.42 40.5493590635045\\
0.43 30.7697004725732\\
0.44 23.7965765961396\\
0.45 18.7141713968253\\
0.46 14.9274258905719\\
0.47 12.0552479833882\\
0.48 9.84398968515933\\
0.49 8.11737464101751\\
0.5 6.75053403557477\\
};
\addplot [
color=blue,
solid,
line width=3.0pt
]
table[row sep=crcr]{
0.01 4586.96579049127\\
0.015 4669.35141336489\\
0.02 4722.93791443017\\
0.025 4752.94522155225\\
0.03 4763.27738723247\\
0.035 4757.67040255542\\
0.04 4739.16056915778\\
0.045 4710.23177530715\\
0.05 4672.89399458139\\
0.055 4628.78628266927\\
0.06 4579.33527834417\\
0.065 4525.59186556739\\
0.07 4468.36050113537\\
0.075 4408.43480105299\\
0.08 4346.2504699863\\
0.085 4282.37867519474\\
0.09 4217.06556365188\\
0.095 4150.64446189884\\
0.1 4083.29021373063\\
0.105 4015.203375057\\
0.11 3946.52632381572\\
0.115 3877.33371775051\\
0.12 3807.66990000492\\
0.125 3737.67504840824\\
0.13 3667.31100336749\\
0.135 3596.67106051344\\
0.14 3525.77625475653\\
0.145 3454.58338209866\\
0.15 3383.16150514192\\
0.155 3311.50594093994\\
0.16 3239.58918792585\\
0.165 3167.50156290352\\
0.17 3095.18229904237\\
0.175 3022.64193264399\\
0.18 2949.93400273844\\
0.185 2877.06270270813\\
0.19 2803.98819973255\\
0.195 2730.79927217762\\
0.2 2657.49285654186\\
0.205 2584.13065671242\\
0.21 2510.70131257459\\
0.215 2437.27107557108\\
0.22 2363.88649189232\\
0.225 2290.61602550676\\
0.23 2217.46270587434\\
0.235 2144.56315772349\\
0.24 2071.93215275807\\
0.245 1999.67271337441\\
0.25 1927.86592006237\\
0.255 1856.56776177061\\
0.26 1785.89685294898\\
0.265 1715.92042915172\\
0.27 1646.76609246792\\
0.275 1578.50054716705\\
0.28 1511.23010226309\\
0.285 1445.05991938775\\
0.29 1380.07234506461\\
0.295 1316.37577566914\\
0.3 1254.06263955344\\
0.305 1193.21232696598\\
0.31 1133.90594528964\\
0.315 1076.22446217155\\
0.32 1020.24717487226\\
0.325 966.005194429354\\
0.33 913.592561189337\\
0.335 863.018088674794\\
0.34 814.341058337135\\
0.345 767.579809046979\\
0.35 722.740137219518\\
0.355 679.840678389015\\
0.36 638.882200736505\\
0.365 599.833597758304\\
0.37 562.691723779679\\
0.375 527.42510990988\\
0.38 493.993567167018\\
0.385 462.359366514156\\
0.39 432.473353625906\\
0.395 404.284216865479\\
0.4 377.724782875468\\
0.405 352.745779842513\\
0.41 329.281314536127\\
0.415 307.268356981236\\
0.42 286.639742702145\\
0.425 267.328317857211\\
0.43 249.267801004658\\
0.435 232.393037866451\\
0.44 216.638286937927\\
0.445 201.940228057254\\
0.45 188.238621818157\\
0.455 175.471603194421\\
0.46 163.581718119076\\
0.465 152.514787073089\\
0.47 142.21743858946\\
0.475 132.639242014624\\
0.48 123.731692554757\\
0.485 115.450808726443\\
0.49 107.753647329142\\
0.495 100.598231201425\\
0.5 93.9483635639049\\
};
\addplot [
color=green,
solid,
line width=3.0pt
]
table[row sep=crcr]{
0.01 4408.93336494782\\
0.015 4638.0500124399\\
0.02 4955.80596540456\\
0.025 5351.67138783141\\
0.03 5816.29287478141\\
0.035 6341.71395505546\\
0.04 6919.99608931029\\
0.045 7544.26163343629\\
0.05 8207.56792627308\\
0.055 8903.59152848384\\
0.06 9626.46816382693\\
0.065 10370.9059446721\\
0.07 11131.8472058422\\
0.075 11905.2154223255\\
0.08 12686.5365248127\\
0.085 13472.5559315932\\
0.09 14259.9190636573\\
0.095 15045.4798538292\\
0.1 15827.0120697751\\
0.105 16601.9974642909\\
0.11 17368.3121365752\\
0.115 18124.1298814298\\
0.12 18867.6550063646\\
0.125 19597.1744163391\\
0.13 20311.241465494\\
0.135 21008.8439787103\\
0.14 21688.0511085081\\
0.145 22348.0256962404\\
0.15 22987.291795476\\
0.155 23604.962045563\\
0.16 24199.3499684654\\
0.165 24770.0267061556\\
0.17 25315.6436358218\\
0.175 25834.6157657642\\
0.18 26326.1885800739\\
0.185 26789.3621523706\\
0.19 27222.39517404\\
0.195 27625.0333967703\\
0.2 27995.5480406781\\
0.205 28333.4122285762\\
0.21 28636.84375241\\
0.215 28905.1320758602\\
0.22 29137.2151964063\\
0.225 29332.3785075954\\
0.23 29489.2000759932\\
0.235 29607.7352334434\\
0.24 29686.5112819401\\
0.245 29725.1323319934\\
0.25 29723.1951208119\\
0.255 29680.3618359394\\
0.26 29596.1641967704\\
0.265 29470.7549006129\\
0.27 29304.4711591611\\
0.275 29097.6677302606\\
0.28 28850.5382515642\\
0.285 28564.1480025576\\
0.29 28239.4463707471\\
0.295 27877.2975416735\\
0.3 27479.8555228234\\
0.305 27048.1463065939\\
0.31 26583.8839143295\\
0.315 26089.6661669415\\
0.32 25567.3699945483\\
0.325 25019.0618377088\\
0.33 24447.8152467852\\
0.335 23855.522597988\\
0.34 23245.2722303229\\
0.345 22619.9507430419\\
0.35 21981.6026948577\\
0.355 21333.4314124905\\
0.36 20678.0621152434\\
0.365 20017.9115291002\\
0.37 19355.7360116301\\
0.375 18693.8842954204\\
0.38 18034.453483068\\
0.385 17379.830492667\\
0.39 16731.8412864198\\
0.395 16092.5564014159\\
0.4 15463.4076054428\\
0.405 14845.9587607965\\
0.41 14241.4147758115\\
0.415 13651.1182098714\\
0.42 13075.8753963428\\
0.425 12516.4877483376\\
0.43 11973.6096199307\\
0.435 11447.69464932\\
0.44 10939.242418844\\
0.445 10448.3177102447\\
0.45 9975.08676287293\\
0.455 9519.6217904966\\
0.46 9081.75085899814\\
0.465 8661.27233053358\\
0.47 8258.14624999457\\
0.475 7871.86456782054\\
0.48 7502.16990616267\\
0.485 7148.58857079386\\
0.49 6810.87869829586\\
0.495 6488.3544377453\\
0.5 6180.64209633605\\
};
\addplot [
color=red,
solid,
line width=3.0pt
]
table[row sep=crcr]{
0.01 17990.0364720629\\
0.015 18614.7502171295\\
0.02 19354.6859818873\\
0.025 20191.4795409618\\
0.03 21106.6052344964\\
0.035 22085.154205089\\
0.04 23112.5356542379\\
0.045 24176.7785700776\\
0.05 25266.4370335051\\
0.055 26371.82053014\\
0.06 27484.7831120809\\
0.065 28598.2477104197\\
0.07 29705.6827233333\\
0.075 30802.7310750836\\
0.08 31883.8967625526\\
0.085 32946.3250924312\\
0.09 33986.3384015723\\
0.095 35000.9890898293\\
0.1 35988.397132972\\
0.105 36946.3644739384\\
0.11 37873.1597578222\\
0.115 38767.2978932628\\
0.12 39627.2335955304\\
0.125 40451.8433404552\\
0.13 41239.7968409756\\
0.135 41990.7847987725\\
0.14 42702.9343284957\\
0.145 43375.6700744144\\
0.15 44007.8817650986\\
0.155 44598.9985802323\\
0.16 45147.236930336\\
0.165 45652.9563818551\\
0.17 46114.6709489237\\
0.175 46530.8303906902\\
0.18 46901.1637294156\\
0.185 47224.8615369621\\
0.19 47499.9376586299\\
0.195 47726.8914845507\\
0.2 47903.8923467081\\
0.205 48030.9574071349\\
0.21 48106.1489222801\\
0.215 48129.2601942557\\
0.22 48099.5437132773\\
0.225 48016.8002261958\\
0.23 47879.6280504361\\
0.235 47689.1309704813\\
0.24 47443.906415212\\
0.245 47144.3427849669\\
0.25 46790.7221875152\\
0.255 46383.2898172779\\
0.26 45922.4457074061\\
0.265 45409.1182136696\\
0.27 44844.7766800988\\
0.275 44230.604368022\\
0.28 43567.797563587\\
0.285 42858.6289843833\\
0.29 42105.107044153\\
0.295 41309.282947069\\
0.3 40474.6949938304\\
0.305 39603.4326686307\\
0.31 38698.4317387349\\
0.315 37763.6898230498\\
0.32 36802.3197115119\\
0.325 35817.3696505437\\
0.33 34813.464123252\\
0.335 33793.2853148833\\
0.34 32761.2406282732\\
0.345 31721.2254041929\\
0.35 30676.0394013118\\
0.355 29629.9208247606\\
0.36 28586.2897696375\\
0.365 27548.0888101422\\
0.37 26518.8254390057\\
0.375 25501.3170419991\\
0.38 24498.0044904506\\
0.385 23511.6615864986\\
0.39 22544.3055372404\\
0.395 21598.1998732682\\
0.4 20674.7282672808\\
0.405 19775.537201153\\
0.41 18901.7152675083\\
0.415 18054.5099357486\\
0.42 17234.4994154661\\
0.425 16442.2373733946\\
0.43 15678.0968582784\\
0.435 14942.212477027\\
0.44 14234.7498085643\\
0.445 13555.3814545893\\
0.45 12903.90677174\\
0.455 12279.9905269503\\
0.46 11683.04331389\\
0.465 11112.458129538\\
0.47 10567.8067971074\\
0.475 10048.145803451\\
0.48 9552.82264569665\\
0.485 9080.97122386753\\
0.49 8631.98814201726\\
0.495 8204.77070306454\\
0.5 7798.61304598581\\
};
\tikzstyle{every node}=[font=\LARGE]
\end{axis}
\begin{scope}[x={(image.south east)},y={(image.north west)},shift={(-18pt,0)}]
\node at (0.62,0.8) {(b)};
\node[anchor=south west,inner sep=0] (mode) at  (0.51,0.03) {\raisebox{0pt}{\includegraphics[scale=0.05]{forward_mode1.png}}};
\end{scope}
\end{tikzpicture}%
}
\subfigure{\definecolor{mycolor1}{rgb}{0,0,0.5625}%
\definecolor{mycolor2}{rgb}{0,0,0.8125}%
\definecolor{mycolor3}{rgb}{0,0.125,1}%
\definecolor{mycolor4}{rgb}{0,0.375,1}%
\definecolor{mycolor5}{rgb}{0,0.6875,1}%
\definecolor{mycolor6}{rgb}{0,0.9375,1}%
\definecolor{mycolor7}{rgb}{0.25,1,0.75}%
\definecolor{mycolor8}{rgb}{0.5,1,0.5}%
\definecolor{mycolor9}{rgb}{0.8125,1,0.1875}%
\definecolor{mycolor10}{rgb}{1,0.875,0}%
\definecolor{mycolor11}{rgb}{1,0.625,0}%
\definecolor{mycolor12}{rgb}{1,0.3125,0}%
\definecolor{mycolor13}{rgb}{1,0.0625,0}%
\begin{tikzpicture}[scale = 0.44*\columnwidth/(3.5in)]
\pgfplotsset{try min ticks=3}
\pgfplotsset{max space between ticks=50pt}
\begin{axis}[%
width=3.5in,
height=2.33in,
xmin=0.2,
xmax=0.35,
xlabel={$a$ ($\mu$m)},
ymin=0.1025,
ymax=0.2525,
ylabel={$b$ ($\mu$m)},
title={},
colorbar horizontal,
colormap/jet,
colorbar style={
at={(0,1.05)},anchor=south west,
xticklabel pos=upper, font = \LARGE, width = 1*
\pgfkeysvalueof{/pgfplots/parent axis width},
xtick = {1,2,3,4,5,6}
},
point meta min=2.20042786519937,
point meta max=6.6873681119492,
xtick = {0.22,0.26,0.30,0.34},
ytick = {0.12,0.16,0.20,0.24}
]
\addplot [solid,fill=mycolor1,draw=black,forget plot] table[row sep=crcr]{
0.2 0.252500167696627\\
0.21 0.252500229879186\\
0.22 0.252500282742584\\
0.23 0.252500326610778\\
0.24 0.252500362462018\\
0.25 0.252500391784445\\
0.26 0.252500415570732\\
0.27 0.252500433892589\\
0.28 0.252500446334496\\
0.29 0.252500452288684\\
0.3 0.252500451559682\\
0.31 0.252500444427132\\
0.32 0.25250043167719\\
0.33 0.252500414324293\\
0.34 0.252500393427384\\
0.35 0.252500369998137\\
0.350000369998137 0.2525\\
0.350000412171012 0.2425\\
0.350000454622354 0.2325\\
0.350000497488267 0.2225\\
0.350000540953824 0.2125\\
0.350000584780131 0.2025\\
0.35000062883597 0.1925\\
0.350000672555163 0.1825\\
0.350000715062387 0.1725\\
0.350000754887508 0.1625\\
0.350000790060917 0.1525\\
0.350000817777111 0.1425\\
0.350000834841016 0.1325\\
0.350000838281016 0.1225\\
0.350000826360032 0.1125\\
0.350000801883076 0.1025\\
0.35 0.102499198116924\\
0.34 0.102499185052031\\
0.33 0.102499175071938\\
0.32 0.102499168990679\\
0.31 0.102499167718643\\
0.3 0.102499172491968\\
0.29 0.102499184901393\\
0.28 0.102499206403742\\
0.27 0.102499239257031\\
0.26 0.102499286307838\\
0.25 0.10249935032693\\
0.24 0.102499434656999\\
0.23 0.102499542477214\\
0.22 0.102499675573809\\
0.21 0.102499831982961\\
0.2 0.1025\\
0.199999888133997 0.1125\\
0.199999761940214 0.1225\\
0.199999636331063 0.1325\\
0.199999528957412 0.1425\\
0.199999455183035 0.1525\\
0.199999422544858 0.1625\\
0.199999428647486 0.1725\\
0.199999462898265 0.1825\\
0.199999512186234 0.1925\\
0.199999566808359 0.2025\\
0.199999622173821 0.2125\\
0.199999677040149 0.2225\\
0.199999730913145 0.2325\\
0.19999978300126 0.2425\\
0.199999832303373 0.2525\\
0.2 0.252500167696627\\
};
\addplot [solid,fill=mycolor2,draw=black,forget plot] table[row sep=crcr]{
0.2 0.252500101031078\\
0.21 0.252500163214052\\
0.22 0.252500216077802\\
0.23 0.252500259946289\\
0.24 0.252500295797768\\
0.25 0.25250032512039\\
0.26 0.252500348906836\\
0.27 0.252500367228815\\
0.28 0.252500379670805\\
0.29 0.252500385625033\\
0.3 0.252500384896026\\
0.31 0.252500377763428\\
0.32 0.252500365013401\\
0.33 0.252500347660389\\
0.34 0.25250032676334\\
0.35 0.252500303333937\\
0.350000303333937 0.2525\\
0.350000345507093 0.2425\\
0.350000387958718 0.2325\\
0.350000430824917 0.2225\\
0.350000474290763 0.2125\\
0.350000518117362 0.2025\\
0.350000562173495 0.1925\\
0.35000060589298 0.1825\\
0.350000648400487 0.1725\\
0.350000688225874 0.1625\\
0.350000723399518 0.1525\\
0.350000751115896 0.1425\\
0.350000768179915 0.1325\\
0.350000771619938 0.1225\\
0.350000759698874 0.1125\\
0.350000735221755 0.1025\\
0.35 0.102499264778245\\
0.34 0.102499251713264\\
0.33 0.102499241733105\\
0.32 0.102499235651805\\
0.31 0.102499234379761\\
0.3 0.102499239153118\\
0.29 0.102499251562626\\
0.28 0.102499273065118\\
0.27 0.102499305918626\\
0.26 0.102499352969747\\
0.25 0.102499416989266\\
0.24 0.102499501319897\\
0.23 0.10249960914083\\
0.22 0.102499742238313\\
0.21 0.102499898648508\\
0.203967784879243 0.1025\\
0.2 0.108459444270555\\
0.199999954799918 0.1125\\
0.199999828605294 0.1225\\
0.199999702995305 0.1325\\
0.199999595620938 0.1425\\
0.199999521846069 0.1525\\
0.199999489207675 0.1625\\
0.199999495310344 0.1725\\
0.199999529561351 0.1825\\
0.199999578849649 0.1925\\
0.199999633472138 0.2025\\
0.199999688837968 0.2125\\
0.199999743704662 0.2225\\
0.199999797578017 0.2325\\
0.19999984966648 0.2425\\
0.199999898968922 0.2525\\
0.2 0.252500101031078\\
};
\addplot [solid,fill=mycolor3,draw=black,forget plot] table[row sep=crcr]{
0.2 0.252500034365529\\
0.21 0.252500096548918\\
0.22 0.25250014941302\\
0.23 0.252500193281799\\
0.24 0.252500229133518\\
0.25 0.252500258456336\\
0.26 0.25250028224294\\
0.27 0.252500300565041\\
0.28 0.252500313007114\\
0.29 0.252500318961381\\
0.3 0.252500318232369\\
0.31 0.252500311099724\\
0.32 0.252500298349612\\
0.33 0.252500280996484\\
0.34 0.252500260099296\\
0.35 0.252500236669737\\
0.350000236669737 0.2525\\
0.350000278843174 0.2425\\
0.350000321295082 0.2325\\
0.350000364161567 0.2225\\
0.350000407627703 0.2125\\
0.350000451454594 0.2025\\
0.350000495511021 0.1925\\
0.350000539230797 0.1825\\
0.350000581738588 0.1725\\
0.35000062156424 0.1625\\
0.350000656738118 0.1525\\
0.350000684454681 0.1425\\
0.350000701518814 0.1325\\
0.350000704958859 0.1225\\
0.350000693037716 0.1125\\
0.350000668560435 0.1025\\
0.35 0.102499331439565\\
0.34 0.102499318374498\\
0.33 0.102499308394272\\
0.32 0.102499302312932\\
0.31 0.102499301040879\\
0.3 0.102499305814268\\
0.29 0.102499318223858\\
0.28 0.102499339726494\\
0.27 0.102499372580221\\
0.26 0.102499419631655\\
0.25 0.102499483651601\\
0.24 0.102499567982795\\
0.23 0.102499675804447\\
0.22 0.102499808902817\\
0.21 0.102499965314054\\
0.207935569758486 0.1025\\
0.201192977038926 0.1125\\
0.2 0.11420098142465\\
0.199999895270374 0.1225\\
0.199999769659547 0.1325\\
0.199999662284464 0.1425\\
0.199999588509104 0.1525\\
0.199999555870492 0.1625\\
0.199999561973201 0.1725\\
0.199999596224437 0.1825\\
0.199999645513063 0.1925\\
0.199999700135916 0.2025\\
0.199999755502116 0.2125\\
0.199999810369176 0.2225\\
0.19999986424289 0.2325\\
0.1999999163317 0.2425\\
0.199999965634471 0.2525\\
0.2 0.252500034365529\\
};
\addplot [solid,fill=mycolor4,draw=black,forget plot] table[row sep=crcr]{
0.2 0.245948694937985\\
0.205194266227953 0.2525\\
0.21 0.252500029883784\\
0.22 0.252500082748239\\
0.23 0.25250012661731\\
0.24 0.252500162469268\\
0.25 0.252500191792281\\
0.26 0.252500215579044\\
0.27 0.252500233901267\\
0.28 0.252500246343423\\
0.29 0.25250025229773\\
0.3 0.252500251568713\\
0.31 0.25250024443602\\
0.32 0.252500231685823\\
0.33 0.25250021433258\\
0.34 0.252500193435253\\
0.35 0.252500170005537\\
0.350000170005537 0.2525\\
0.350000212179255 0.2425\\
0.350000254631447 0.2325\\
0.350000297498217 0.2225\\
0.350000340964643 0.2125\\
0.350000384791826 0.2025\\
0.350000428848547 0.1925\\
0.350000472568614 0.1825\\
0.350000515076688 0.1725\\
0.350000554902606 0.1625\\
0.350000590076719 0.1525\\
0.350000617793466 0.1425\\
0.350000634857712 0.1325\\
0.350000638297781 0.1225\\
0.350000626376559 0.1125\\
0.350000601899114 0.1025\\
0.35 0.102499398100886\\
0.34 0.102499385035732\\
0.33 0.102499375055439\\
0.32 0.102499368974059\\
0.31 0.102499367701997\\
0.3 0.102499372475418\\
0.29 0.102499384885091\\
0.28 0.10249940638787\\
0.27 0.102499439241816\\
0.26 0.102499486293564\\
0.25 0.102499550313937\\
0.24 0.102499634645693\\
0.23 0.102499742468063\\
0.22 0.10249987556732\\
0.212044545537899 0.1025\\
0.21 0.105083534142031\\
0.20489797606686 0.1125\\
0.2 0.119483676999862\\
0.199999961935453 0.1225\\
0.19999983632379 0.1325\\
0.199999728947991 0.1425\\
0.199999655172139 0.1525\\
0.199999622533309 0.1625\\
0.199999628636059 0.1725\\
0.199999662887523 0.1825\\
0.199999712176478 0.1925\\
0.199999766799695 0.2025\\
0.199999822166264 0.2125\\
0.199999877033689 0.2225\\
0.199999930907763 0.2325\\
0.19999998299692 0.2425\\
0.2 0.245948694937985\\
};
\addplot [solid,fill=mycolor5,draw=black,forget plot] table[row sep=crcr]{
0.2 0.232966001078362\\
0.207262761352455 0.2425\\
0.21 0.245872530204704\\
0.216957613773048 0.2525\\
0.22 0.252500016083457\\
0.23 0.252500059952821\\
0.24 0.252500095805017\\
0.25 0.252500125128226\\
0.26 0.252500148915147\\
0.27 0.252500167237493\\
0.28 0.252500179679732\\
0.29 0.252500185634078\\
0.3 0.252500184905057\\
0.31 0.252500177772317\\
0.32 0.252500165022034\\
0.33 0.252500147668675\\
0.34 0.252500126771209\\
0.35 0.252500103341337\\
0.350000103341337 0.2525\\
0.350000145515336 0.2425\\
0.350000187967811 0.2325\\
0.350000230834867 0.2225\\
0.350000274301583 0.2125\\
0.350000318129058 0.2025\\
0.350000362186072 0.1925\\
0.350000405906431 0.1825\\
0.350000448414789 0.1725\\
0.350000488240971 0.1625\\
0.350000523415319 0.1525\\
0.350000551132252 0.1425\\
0.350000568196611 0.1325\\
0.350000571636703 0.1225\\
0.350000559715401 0.1125\\
0.350000535237793 0.1025\\
0.35 0.102499464762207\\
0.34 0.102499451696965\\
0.33 0.102499441716606\\
0.32 0.102499435635185\\
0.31 0.102499434363115\\
0.3 0.102499439136568\\
0.29 0.102499451546324\\
0.28 0.102499473049246\\
0.27 0.102499505903411\\
0.26 0.102499552955473\\
0.25 0.102499616976272\\
0.24 0.10249970130859\\
0.23 0.10249980913168\\
0.22 0.102499942231824\\
0.216306660810267 0.1025\\
0.21 0.110469239727613\\
0.208602975094794 0.1125\\
0.201532719857136 0.1225\\
0.2 0.124776863777514\\
0.199999902988032 0.1325\\
0.199999795611517 0.1425\\
0.199999721835173 0.1525\\
0.199999689196126 0.1625\\
0.199999695298917 0.1725\\
0.199999729550609 0.1825\\
0.199999778839892 0.1925\\
0.199999833463474 0.2025\\
0.199999888830412 0.2125\\
0.199999943698203 0.2225\\
0.199999997572636 0.2325\\
0.2 0.232966001078362\\
};
\addplot [solid,fill=mycolor6,draw=black,forget plot] table[row sep=crcr]{
0.2 0.220611350743653\\
0.201249087743351 0.2225\\
0.208566208814771 0.2325\\
0.21 0.234331676138707\\
0.218372140693888 0.2425\\
0.22 0.244056297670201\\
0.23 0.251443451397381\\
0.231872020361052 0.2525\\
0.24 0.252500029140767\\
0.25 0.252500058464171\\
0.26 0.252500082251251\\
0.27 0.252500100573719\\
0.28 0.252500113016041\\
0.29 0.252500118970427\\
0.3 0.2525001182414\\
0.31 0.252500111108613\\
0.32 0.252500098358246\\
0.33 0.252500081004771\\
0.34 0.252500060107165\\
0.35 0.252500036677137\\
0.350000036677137 0.2525\\
0.350000078851417 0.2425\\
0.350000121304175 0.2325\\
0.350000164171517 0.2225\\
0.350000207638522 0.2125\\
0.35000025146629 0.2025\\
0.350000295523598 0.1925\\
0.350000339244248 0.1825\\
0.350000381752889 0.1725\\
0.350000421579337 0.1625\\
0.350000456753919 0.1525\\
0.350000484471037 0.1425\\
0.35000050153551 0.1325\\
0.350000504975625 0.1225\\
0.350000493054244 0.1125\\
0.350000468576472 0.1025\\
0.35 0.102499531423528\\
0.34 0.102499518358199\\
0.33 0.102499508377773\\
0.32 0.102499502296312\\
0.31 0.102499501024233\\
0.3 0.102499505797718\\
0.29 0.102499518207556\\
0.28 0.102499539710622\\
0.27 0.102499572565006\\
0.26 0.102499619617382\\
0.25 0.102499683638608\\
0.24 0.102499767971488\\
0.23 0.102499875795296\\
0.220668380804678 0.1025\\
0.22 0.103211706345912\\
0.212634480493373 0.1125\\
0.21 0.115625644181969\\
0.20510534180658 0.1225\\
0.2 0.13008401333232\\
0.199999969652274 0.1325\\
0.199999862275044 0.1425\\
0.199999788498208 0.1525\\
0.199999755858943 0.1625\\
0.199999761961774 0.1725\\
0.199999796213695 0.1825\\
0.199999845503307 0.1925\\
0.199999900127253 0.2025\\
0.19999995549456 0.2125\\
0.2 0.220611350743653\\
};
\addplot [solid,fill=mycolor7,draw=black,forget plot] table[row sep=crcr]{
0.2 0.208497911525674\\
0.202372525714686 0.2125\\
0.209284608827376 0.2225\\
0.21 0.22345962306968\\
0.218921469739622 0.2325\\
0.22 0.233554486317419\\
0.23 0.241074605549311\\
0.232504189761967 0.2425\\
0.24 0.246900526733612\\
0.25 0.251332436345528\\
0.253447172166881 0.2525\\
0.26 0.252500015587355\\
0.27 0.252500033909945\\
0.28 0.25250004635235\\
0.29 0.252500052306775\\
0.3 0.252500051577744\\
0.31 0.252500044444909\\
0.32 0.252500031694457\\
0.33 0.252500014340866\\
0.336862403338692 0.2525\\
0.34 0.251055290345211\\
0.35 0.245389783420273\\
0.350000012187498 0.2425\\
0.350000054640539 0.2325\\
0.350000097508167 0.2225\\
0.350000140975462 0.2125\\
0.350000184803522 0.2025\\
0.350000228861123 0.1925\\
0.350000272582065 0.1825\\
0.350000315090989 0.1725\\
0.350000354917703 0.1625\\
0.35000039009252 0.1525\\
0.350000417809822 0.1425\\
0.350000434874409 0.1325\\
0.350000438314547 0.1225\\
0.350000426393086 0.1125\\
0.350000401915152 0.1025\\
0.35 0.102499598084849\\
0.34 0.102499585019433\\
0.33 0.10249957503894\\
0.32 0.102499568957438\\
0.31 0.102499567685351\\
0.3 0.102499572458868\\
0.29 0.102499584868789\\
0.28 0.102499606371998\\
0.27 0.102499639226601\\
0.26 0.10249968627929\\
0.25 0.102499750300943\\
0.24 0.102499834634386\\
0.23 0.102499942458913\\
0.225676882607662 0.1025\\
0.22 0.108544867459676\\
0.216863621353721 0.1125\\
0.21 0.120643251850019\\
0.208677963756024 0.1225\\
0.201972455410639 0.1325\\
0.2 0.135882096551133\\
0.19999992893857 0.1425\\
0.199999855161242 0.1525\\
0.19999982252176 0.1625\\
0.199999828624632 0.1725\\
0.199999862876781 0.1825\\
0.199999912166722 0.1925\\
0.199999966791031 0.2025\\
0.2 0.208497911525674\\
};
\addplot [solid,fill=mycolor8,draw=black,forget plot] table[row sep=crcr]{
0.2 0.196375526049483\\
0.203134092922628 0.2025\\
0.209510234695448 0.2125\\
0.21 0.213200463843453\\
0.218683982945418 0.2225\\
0.22 0.223831714387672\\
0.23 0.231521066028018\\
0.231683906702283 0.2325\\
0.24 0.237457489716841\\
0.25 0.241898130018458\\
0.251789531005527 0.2425\\
0.26 0.245438788447447\\
0.27 0.247993424032119\\
0.28 0.249637365036322\\
0.29 0.250373548116168\\
0.3 0.250107015063242\\
0.31 0.248682336414373\\
0.32 0.24596077738812\\
0.328540523816016 0.2425\\
0.33 0.241849789334305\\
0.34 0.23635346088258\\
0.345920770020356 0.2325\\
0.35 0.229695324237055\\
0.350000030844816 0.2225\\
0.350000074312402 0.2125\\
0.350000118140754 0.2025\\
0.350000162198649 0.1925\\
0.350000205919882 0.1825\\
0.35000024842909 0.1725\\
0.350000288256069 0.1625\\
0.35000032343112 0.1525\\
0.350000351148607 0.1425\\
0.350000368213308 0.1325\\
0.350000371653469 0.1225\\
0.350000359731928 0.1125\\
0.350000335253831 0.1025\\
0.35 0.102499664746169\\
0.34 0.102499651680667\\
0.33 0.102499641700108\\
0.32 0.102499635618565\\
0.31 0.102499634346469\\
0.3 0.102499639120018\\
0.29 0.102499651530022\\
0.28 0.102499673033374\\
0.27 0.102499705888196\\
0.26 0.102499752941199\\
0.25 0.102499816963279\\
0.24 0.102499901297283\\
0.2308460392567 0.1025\\
0.23 0.103260943542472\\
0.221345849211177 0.1125\\
0.22 0.113848556390258\\
0.212753769794906 0.1225\\
0.21 0.125910572114697\\
0.205593183907489 0.1325\\
0.2 0.142090426177105\\
0.199999995602096 0.1425\\
0.199999921824277 0.1525\\
0.199999889184577 0.1625\\
0.19999989528749 0.1725\\
0.199999929539867 0.1825\\
0.199999978830136 0.1925\\
0.2 0.196375526049483\\
};
\addplot [solid,fill=mycolor9,draw=black,forget plot] table[row sep=crcr]{
0.2 0.18327034256599\\
0.203743983259529 0.1925\\
0.209379248007307 0.2025\\
0.21 0.203464292048811\\
0.217838614494397 0.2125\\
0.22 0.214795533617618\\
0.229740020413129 0.2225\\
0.23 0.222702879877878\\
0.24 0.228754859497324\\
0.248412483860121 0.2325\\
0.25 0.233238256796977\\
0.26 0.236671525945479\\
0.27 0.238996478200425\\
0.28 0.240293314123014\\
0.29 0.240504643028396\\
0.3 0.23949977031267\\
0.31 0.237138163701474\\
0.32 0.23334945261185\\
0.32174239335189 0.2325\\
0.33 0.228099846419505\\
0.338723122320307 0.2225\\
0.34 0.221630205863273\\
0.35 0.214259775164737\\
0.350000007649341 0.2125\\
0.350000051477986 0.2025\\
0.350000095536174 0.1925\\
0.350000139257699 0.1825\\
0.35000018176719 0.1725\\
0.350000221594435 0.1625\\
0.350000256769721 0.1525\\
0.350000284487392 0.1425\\
0.350000301552207 0.1325\\
0.350000304992391 0.1225\\
0.350000293070771 0.1125\\
0.35000026859251 0.1025\\
0.35 0.10249973140749\\
0.34 0.1024997183419\\
0.33 0.102499708361275\\
0.32 0.102499702279692\\
0.31 0.102499701007588\\
0.3 0.102499705781168\\
0.29 0.102499718191254\\
0.28 0.10249973969475\\
0.27 0.102499772549791\\
0.26 0.102499819603108\\
0.25 0.102499883625614\\
0.24 0.102499967960181\\
0.237028539291366 0.1025\\
0.23 0.108821599789158\\
0.226554471743667 0.1125\\
0.22 0.119067656080106\\
0.217125155603095 0.1225\\
0.21 0.131324578240978\\
0.209213912404338 0.1325\\
0.203550833202995 0.1425\\
0.2 0.150939546779066\\
0.199999988487311 0.1525\\
0.199999955847394 0.1625\\
0.199999961950347 0.1725\\
0.199999996202953 0.1825\\
0.2 0.18327034256599\\
};
\addplot [solid,fill=mycolor10,draw=black,forget plot] table[row sep=crcr]{
0.35 0.199053503913693\\
0.3500000288737 0.1925\\
0.350000072595516 0.1825\\
0.350000115105291 0.1725\\
0.350000154932801 0.1625\\
0.350000190108321 0.1525\\
0.350000217826178 0.1425\\
0.350000234891106 0.1325\\
0.350000238331313 0.1225\\
0.350000226409613 0.1125\\
0.350000201931189 0.1025\\
0.35 0.102499798068811\\
0.34 0.102499785003134\\
0.33 0.102499775022442\\
0.32 0.102499768940818\\
0.31 0.102499767668706\\
0.3 0.102499772442318\\
0.29 0.102499784852487\\
0.28 0.102499806356126\\
0.27 0.102499839211386\\
0.26 0.102499886265017\\
0.25 0.10249995028795\\
0.244105395598546 0.1025\\
0.24 0.105601879349722\\
0.232266442024375 0.1125\\
0.23 0.114414961844239\\
0.221932551957703 0.1225\\
0.22 0.124516662388521\\
0.213661546211036 0.1325\\
0.21 0.137792272259538\\
0.207352466470486 0.1425\\
0.203349891795883 0.1525\\
0.201453014989049 0.1625\\
0.201961823397386 0.1725\\
0.204652559462688 0.1825\\
0.20923018304243 0.1925\\
0.21 0.193848024875118\\
0.216725222620202 0.2025\\
0.22 0.206211012846843\\
0.227508538496381 0.2125\\
0.23 0.214515384431017\\
0.24 0.220568412583598\\
0.244352725374063 0.2225\\
0.25 0.225082800010065\\
0.26 0.228277399962399\\
0.27 0.230281330868218\\
0.28 0.231093706683391\\
0.29 0.230605590903501\\
0.3 0.22869382669446\\
0.31 0.225290767128912\\
0.315931944893872 0.2225\\
0.32 0.220385697411488\\
0.33 0.21415958116506\\
0.332384156648145 0.2125\\
0.34 0.206927579325782\\
0.345690375578821 0.2025\\
0.35 0.199053503913693\\
};
\addplot [solid,fill=mycolor11,draw=black,forget plot] table[row sep=crcr]{
0.35 0.183857060777063\\
0.350000005933333 0.1825\\
0.350000048443391 0.1725\\
0.350000088271167 0.1625\\
0.350000123446921 0.1525\\
0.350000151164963 0.1425\\
0.350000168230005 0.1325\\
0.350000171670234 0.1225\\
0.350000159748456 0.1125\\
0.350000135269868 0.1025\\
0.35 0.102499864730132\\
0.34 0.102499851664368\\
0.33 0.102499841683609\\
0.32 0.102499835601945\\
0.31 0.102499834329824\\
0.3 0.102499839103468\\
0.29 0.10249985151372\\
0.28 0.102499873017502\\
0.27 0.102499905872981\\
0.26 0.102499952926925\\
0.252647503280267 0.1025\\
0.25 0.104159840281216\\
0.24 0.111574203958679\\
0.238962080003795 0.1125\\
0.23 0.120072239248882\\
0.227577521176593 0.1225\\
0.22 0.130407317520826\\
0.218338493016987 0.1325\\
0.211554905299956 0.1425\\
0.21 0.145709387507673\\
0.207399076783572 0.1525\\
0.205756043157935 0.1625\\
0.206532466699534 0.1725\\
0.209586128970411 0.1825\\
0.21 0.183389057129595\\
0.215858372206155 0.1925\\
0.22 0.197697987956581\\
0.225251804533297 0.2025\\
0.23 0.206525480123515\\
0.23967221909701 0.2125\\
0.24 0.212700872893535\\
0.25 0.217090233562258\\
0.26 0.220011768499376\\
0.27 0.22159312893784\\
0.28 0.221758862666891\\
0.29 0.220389899346278\\
0.3 0.217446612917188\\
0.31 0.212979583062818\\
0.31087275400768 0.2125\\
0.32 0.20700437568046\\
0.326524990309931 0.2025\\
0.33 0.199949132934779\\
0.339562980405226 0.1925\\
0.34 0.192144756455586\\
0.35 0.183857060777063\\
};
\addplot [solid,fill=mycolor12,draw=black,forget plot] table[row sep=crcr]{
0.35 0.167925718011397\\
0.350000021609533 0.1625\\
0.350000056785522 0.1525\\
0.350000084503748 0.1425\\
0.350000101568904 0.1325\\
0.350000105009156 0.1225\\
0.350000093087298 0.1125\\
0.350000068608548 0.1025\\
0.35 0.102499931391452\\
0.34 0.102499918325601\\
0.33 0.102499908344776\\
0.32 0.102499902263071\\
0.31 0.102499900990942\\
0.3 0.102499905764618\\
0.29 0.102499918174953\\
0.28 0.102499939678878\\
0.27 0.102499972534576\\
0.264163019687478 0.1025\\
0.26 0.104622732999586\\
0.25 0.110687684925318\\
0.247526947019506 0.1125\\
0.24 0.117797645868437\\
0.234321767980385 0.1225\\
0.23 0.126298524465358\\
0.224089122016105 0.1325\\
0.22 0.137484243734593\\
0.216676802066409 0.1425\\
0.212010779624237 0.1525\\
0.210082671675463 0.1625\\
0.211511739445647 0.1725\\
0.21591852375025 0.1825\\
0.22 0.188669526196826\\
0.223568449207139 0.1925\\
0.23 0.198394547294256\\
0.236294617545925 0.2025\\
0.24 0.204792499150661\\
0.25 0.209014063183138\\
0.26 0.21162889750333\\
0.26873794172718 0.2125\\
0.27 0.212626917056341\\
0.272182768603738 0.2125\\
0.28 0.211965860831191\\
0.29 0.209567061946959\\
0.3 0.205535943787445\\
0.305691567877465 0.2025\\
0.31 0.199937389134632\\
0.32 0.193097867656731\\
0.320806301413963 0.1925\\
0.33 0.185268187307408\\
0.333396154149838 0.1825\\
0.34 0.176822332105183\\
0.345013502531428 0.1725\\
0.35 0.167925718011397\\
};
\addplot [solid,fill=mycolor13,draw=black,forget plot] table[row sep=crcr]{
0.35 0.148937075630099\\
0.350000017842533 0.1425\\
0.350000034907803 0.1325\\
0.350000038348078 0.1225\\
0.35000002642614 0.1125\\
0.350000001947227 0.1025\\
0.35 0.102499998052773\\
0.34 0.102499984986835\\
0.33 0.102499975005943\\
0.32 0.102499968924198\\
0.31 0.10249996765206\\
0.3 0.102499972425768\\
0.29 0.102499984836185\\
0.282948392997508 0.1025\\
0.28 0.10336273404087\\
0.27 0.107253346516255\\
0.26 0.111846513332394\\
0.258887064785463 0.1125\\
0.25 0.117616747126177\\
0.242660975999966 0.1225\\
0.24 0.124430151193699\\
0.23061101747712 0.1325\\
0.23 0.133124253404938\\
0.222558368543819 0.1425\\
0.22 0.14704740366126\\
0.217632704354682 0.1525\\
0.216104858198707 0.1625\\
0.21777550440227 0.1725\\
0.22 0.177409432407634\\
0.223375883109625 0.1825\\
0.23 0.189614534053189\\
0.233946456996179 0.1925\\
0.24 0.19637622176059\\
0.25 0.200542095671894\\
0.258867903310643 0.2025\\
0.26 0.202740396430767\\
0.27 0.202960049611896\\
0.273187304469796 0.2025\\
0.28 0.20131151842785\\
0.29 0.197784371801749\\
0.3 0.192684950142062\\
0.300302948480019 0.1925\\
0.31 0.185911966940071\\
0.314564325667129 0.1825\\
0.32 0.178052133702132\\
0.326569663680738 0.1725\\
0.33 0.169311723509644\\
0.337364460063767 0.1625\\
0.34 0.159733314158157\\
0.347107360464222 0.1525\\
0.35 0.148937075630099\\
};
\addplot [solid,fill=red!75!black,draw=black,forget plot] table[row sep=crcr]{
0.33 0.148573536768628\\
0.334471573937055 0.1425\\
0.338814032621344 0.1325\\
0.338239871643226 0.1225\\
0.33 0.114777088656976\\
0.325796653902038 0.1125\\
0.32 0.110812341533126\\
0.31 0.10941602642118\\
0.3 0.109362248811793\\
0.29 0.110422370846586\\
0.28 0.112433511912012\\
0.279795436857756 0.1125\\
0.27 0.115776083080239\\
0.26 0.119867007144134\\
0.254886112573853 0.1225\\
0.25 0.125317019813965\\
0.24 0.132151848226464\\
0.239594937774486 0.1325\\
0.23 0.142302784372255\\
0.229843467823687 0.1425\\
0.224777400478417 0.1525\\
0.223179934290996 0.1625\\
0.225974632570678 0.1725\\
0.23 0.178602855629942\\
0.23404684588357 0.1825\\
0.24 0.186757532229196\\
0.25 0.191038842256178\\
0.258507986236719 0.1925\\
0.26 0.192728619734055\\
0.264065135529699 0.1925\\
0.27 0.192073833116165\\
0.28 0.189181778105031\\
0.29 0.184491146293781\\
0.293340728318706 0.1825\\
0.3 0.17790129688634\\
0.30703163921402 0.1725\\
0.31 0.16985588750558\\
0.318016174422997 0.1625\\
0.32 0.160317290221422\\
0.327219989970533 0.1525\\
0.33 0.148573536768628\\
};
\addplot [solid,fill=red!50!black,draw=black,forget plot] table[row sep=crcr]{
0.24 0.173810657870193\\
0.25 0.178782452220655\\
0.26 0.180153409774197\\
0.27 0.1783991256059\\
0.28 0.174309323977316\\
0.283136065122264 0.1725\\
0.29 0.167474810258895\\
0.296030788196248 0.1625\\
0.3 0.158137616105179\\
0.305037495411574 0.1525\\
0.31 0.143839055561886\\
0.310792850335829 0.1425\\
0.311151810538492 0.1325\\
0.31 0.131115936030406\\
0.3 0.12453816487916\\
0.29 0.122802453789268\\
0.28 0.123686372580784\\
0.27 0.126163980468764\\
0.26 0.129791476090015\\
0.254534854844081 0.1325\\
0.25 0.135413059207715\\
0.241080826768219 0.1425\\
0.24 0.143850678266388\\
0.234831498187267 0.1525\\
0.233595791463177 0.1625\\
0.238636629153676 0.1725\\
0.24 0.173810657870193\\
};
\tikzstyle{every node}=[font=\LARGE]
\end{axis}
\begin{scope}[x={(image.south east)},y={(image.north west)},shift={(-18pt,0)}]
\node[scale=0.8] at (0.62,1) {Gain ($\times 10^{4} \, \text{W}^{-1}\text{m}^{-1}$)};
\node at (0.62,1.15) {{(c)}};
\node at (1.2,1.05) {\textcolor{white}{la}};
\hspace{100pt}
\end{scope}
\end{tikzpicture}%
}
\subfigure{\begin{tikzpicture}[scale = 0.44*\columnwidth/(3.5in)]
\pgfplotsset{try min ticks=3}
\pgfplotsset{max space between ticks=50pt}
\begin{axis}[%
width=3.5in,
height=2.33in,
xmin=0,
xmax=0.5,
xlabel={$\bar{a}$ ($\mu$m)},
ymode=log,
ymin=0.5,
ymax=50000,
yminorticks=false,
legend style={fill=none,draw=none,legend cell align=left,at={(0.5,-0.41)},anchor=north, legend columns=-1},
ylabel={Gain ($\text{W}^{-1}\text{m}^{-1}$)},
ytick = {1,100,10000},
extra y ticks = {10,1000},
extra y tick labels={}
]
\addplot [
color=blue,
loosely dotted,
line width=3.0pt
]
table[row sep=crcr]{
0.01 1.63759657309751\\
0.02 4.59854852766764\\
0.03 7.67673164510369\\
0.04 10.5410971833564\\
0.05 13.1301327682198\\
0.06 15.4596567785426\\
0.07 17.5693586592114\\
0.08 19.4973326747829\\
0.09 21.2797883193683\\
0.1 22.9470003303783\\
0.11 24.524446959218\\
0.12 26.0339184721163\\
0.13 27.4951543382957\\
0.14 28.9243333912679\\
0.15 30.337118531367\\
0.16 31.7476757644258\\
0.17 33.1712532265414\\
0.18 34.6207018654054\\
0.19 36.1095698286214\\
0.2 37.6531145775138\\
0.21 39.2624487747497\\
0.22 40.9503869534832\\
0.23 42.7207400573568\\
0.24 44.5718767403308\\
0.25 46.4673410285737\\
0.26 48.3272349271695\\
0.27 49.9725233033616\\
0.28 51.0385922748695\\
0.29 50.8657428942601\\
0.3 48.4626855801244\\
0.31 42.8602229912452\\
0.32 34.1364851674226\\
0.33 24.2065156551879\\
0.34 15.6068217981789\\
0.35 9.53619808510357\\
0.36 5.74194222418947\\
0.37 3.49430896363743\\
0.38 2.17676153700325\\
0.39 1.39471307192149\\
0.4 0.919683120184542\\
0.41 0.623298905445324\\
0.42 0.433287615419696\\
0.43 0.308194149573319\\
0.44 0.223831137505943\\
0.45 0.165629090834619\\
0.46 0.124646723698741\\
0.47 0.0952469017861365\\
0.48 0.0737898958762664\\
0.49 0.0578853985799927\\
0.5 0.0459292948426317\\
};
\addplot [
color=green,
loosely dotted,
line width=3.0pt
]
table[row sep=crcr]{
0.01 6.29412386829055\\
0.02 19.8793423445545\\
0.03 36.128768063591\\
0.04 52.7931010866928\\
0.05 68.9355464745638\\
0.06 84.3992620855056\\
0.07 98.9581393486971\\
0.08 112.820621757349\\
0.09 126.154281435995\\
0.1 138.830330984163\\
0.11 151.116655783152\\
0.12 163.153413036062\\
0.13 175.026178239836\\
0.14 187.070894763195\\
0.15 198.914610557226\\
0.16 210.985328440606\\
0.17 223.573276868465\\
0.18 236.683201292589\\
0.19 250.387413814692\\
0.2 264.910742456903\\
0.21 280.530159073182\\
0.22 297.598177987848\\
0.23 316.117741672908\\
0.24 337.06226186848\\
0.25 359.825238431682\\
0.26 384.963001128155\\
0.27 411.906159651964\\
0.28 438.541852676307\\
0.29 460.740450877104\\
0.3 469.451298537442\\
0.31 452.39744860678\\
0.32 400.570077362682\\
0.33 321.408254426253\\
0.34 237.139968480907\\
0.35 166.421326698373\\
0.36 114.650721019121\\
0.37 79.1926793663838\\
0.38 55.4754249464746\\
0.39 39.5960470908672\\
0.4 28.785434417921\\
0.41 21.3247347214667\\
0.42 16.0718825596045\\
0.43 12.3122250563187\\
0.44 9.56479971314987\\
0.45 7.52864571089663\\
0.46 5.99482831463565\\
0.47 4.82175597975362\\
0.48 3.92060562461056\\
0.49 3.21408908707035\\
0.5 2.65697862518199\\
};
\addplot [
color=red,
loosely dotted,
line width=3.0pt
]
table[row sep=crcr]{
0.01 14.3526972136132\\
0.02 43.600246424774\\
0.03 77.113207951225\\
0.04 110.514584413683\\
0.05 142.23653190953\\
0.06 172.102496411237\\
0.07 199.921287763872\\
0.08 226.119898867817\\
0.09 251.059093847809\\
0.1 274.662045088553\\
0.11 297.395813688737\\
0.12 319.533373311194\\
0.13 341.263853586601\\
0.14 363.112880772135\\
0.15 384.615767631561\\
0.16 406.41921374491\\
0.17 428.979320792954\\
0.18 452.34686544596\\
0.19 476.669345498021\\
0.2 502.310842378817\\
0.21 529.690687880916\\
0.22 559.335888356911\\
0.23 591.258651849185\\
0.24 626.774894827103\\
0.25 664.905201463845\\
0.26 706.084645043402\\
0.27 748.821117335506\\
0.28 788.796479375885\\
0.29 817.781995005231\\
0.3 819.581819169307\\
0.31 773.752594553235\\
0.32 668.578788366388\\
0.33 522.025358680603\\
0.34 374.418499880082\\
0.35 255.632533085489\\
0.36 171.70807286083\\
0.37 115.957016244913\\
0.38 79.6300636577998\\
0.39 55.8534819154222\\
0.4 39.9955943280976\\
0.41 29.2395859474545\\
0.42 21.7829529176374\\
0.43 16.516343091487\\
0.44 12.7149937808549\\
0.45 9.92762452294534\\
0.46 7.84833097051126\\
0.47 6.27237341130689\\
0.48 5.07012983863776\\
0.49 4.13464204126298\\
0.5 3.40157280297368\\
};
\addplot [
color=blue,
solid,
line width=3.0pt
]
table[row sep=crcr]{
0.01 8.95249087482261\\
0.015 18.8979535544091\\
0.02 31.4321523479918\\
0.025 45.8577103850411\\
0.03 61.5770731129911\\
0.035 78.0957386047877\\
0.04 95.0184190123737\\
0.045 112.037881900124\\
0.05 128.908805135309\\
0.055 145.454788451152\\
0.06 161.543775073582\\
0.065 177.079593431391\\
0.07 191.994775290347\\
0.075 206.250417200766\\
0.08 219.815506044054\\
0.085 232.677663001664\\
0.09 244.834210997467\\
0.095 256.286077477233\\
0.1 267.040918327534\\
0.105 277.108432709292\\
0.11 286.504246794305\\
0.115 295.238386693763\\
0.12 303.324762562634\\
0.125 310.781107335275\\
0.13 317.617896763442\\
0.135 323.852776211464\\
0.14 329.496538134129\\
0.145 334.559014471752\\
0.15 339.055143776502\\
0.155 342.995062817562\\
0.16 346.384047016191\\
0.165 349.240720918351\\
0.17 351.566644574608\\
0.175 353.368482248124\\
0.18 354.659212793842\\
0.185 355.445423172078\\
0.19 355.726695045093\\
0.195 355.51728486567\\
0.2 354.824803406364\\
0.205 353.654681046324\\
0.21 352.011348260329\\
0.215 349.904046374121\\
0.22 347.343654536177\\
0.225 344.3383791527\\
0.23 340.894179776811\\
0.235 337.031239778519\\
0.24 332.753170529468\\
0.245 328.077691572154\\
0.25 323.020302974436\\
0.255 317.592669552475\\
0.26 311.816040461793\\
0.265 305.706923318088\\
0.27 299.290169615021\\
0.275 292.583264655071\\
0.28 285.611392971831\\
0.285 278.399247894079\\
0.29 270.970190052374\\
0.295 263.353585926022\\
0.3 255.577781111656\\
0.305 247.669079636469\\
0.31 239.655277850257\\
0.315 231.565080181251\\
0.32 223.431071695534\\
0.325 215.275545694899\\
0.33 207.132712830901\\
0.335 199.023603994325\\
0.34 190.978093948005\\
0.345 183.02005880243\\
0.35 175.169870417284\\
0.355 167.451458625858\\
0.36 159.884890779441\\
0.365 152.485026976313\\
0.37 145.269456155523\\
0.375 138.252358146061\\
0.38 131.443112022705\\
0.385 124.853143826172\\
0.39 118.4894200098\\
0.395 112.357629105609\\
0.4 106.460137954058\\
0.405 100.800475990341\\
0.41 95.3791722473739\\
0.415 90.1953563969536\\
0.42 85.2467381209326\\
0.425 80.5301053661942\\
0.43 76.0408920320459\\
0.435 71.7741732440457\\
0.44 67.7239521111022\\
0.445 63.8834784600796\\
0.45 60.2464387271804\\
0.455 56.8049544600506\\
0.46 53.5518505970882\\
0.465 50.4790421317228\\
0.47 47.5788995791779\\
0.475 44.8436438699763\\
0.48 42.2650640619842\\
0.485 39.8362203295826\\
0.49 37.5489413562297\\
0.495 35.3958972265648\\
0.5 33.3702429770222\\
};
\addplot [
color=green,
solid,
line width=3.0pt
]
table[row sep=crcr]{
0.01 63.7523368329803\\
0.015 164.11901947372\\
0.02 320.91504165684\\
0.025 537.12039298183\\
0.03 811.537161575696\\
0.035 1139.7294685823\\
0.04 1517.19973614603\\
0.045 1937.36642301055\\
0.05 2393.10149712179\\
0.055 2879.15034996925\\
0.06 3388.89596738437\\
0.065 3916.79004902123\\
0.07 4457.91221951267\\
0.075 5007.34467485599\\
0.08 5561.82220801077\\
0.085 6117.20068039757\\
0.09 6671.61455457176\\
0.095 7221.8182327571\\
0.1 7765.80693520547\\
0.105 8301.00480602864\\
0.11 8827.22610330874\\
0.115 9341.80122957671\\
0.12 9843.94677853141\\
0.125 10333.1952683466\\
0.13 10807.40377453\\
0.135 11266.0121743378\\
0.14 11709.6580419383\\
0.145 12135.2617655325\\
0.15 12544.1806151762\\
0.155 12935.2000170427\\
0.16 13307.5470026294\\
0.165 13661.4987005763\\
0.17 13994.4409540783\\
0.175 14307.2091625976\\
0.18 14600.518597769\\
0.185 14872.5080323113\\
0.19 15123.1170244911\\
0.195 15350.5790428835\\
0.2 15556.6123368556\\
0.205 15738.2992963981\\
0.21 15897.6803167488\\
0.215 16033.1984326544\\
0.22 16145.071519929\\
0.225 16231.362259334\\
0.23 16294.0399778995\\
0.235 16331.5899022025\\
0.24 16344.0382258437\\
0.245 16331.9299223434\\
0.25 16294.7820619117\\
0.255 16232.3941979127\\
0.26 16145.6238415896\\
0.265 16034.2520348943\\
0.27 15900.0785521404\\
0.275 15740.9546178954\\
0.28 15559.7186365977\\
0.285 15356.6976510141\\
0.29 15132.3158839177\\
0.295 14888.008109585\\
0.3 14624.7032561693\\
0.305 14343.9215274975\\
0.31 14047.3109373854\\
0.315 13733.839824803\\
0.32 13407.9813101577\\
0.325 13069.6837918415\\
0.33 12721.1577458111\\
0.335 12363.668852384\\
0.34 11999.0742330044\\
0.345 11627.9629512715\\
0.35 11253.4881539162\\
0.355 10875.9152747972\\
0.36 10497.8783469587\\
0.365 10119.9222361785\\
0.37 9743.55056873448\\
0.375 9370.37874867477\\
0.38 9001.37707260411\\
0.385 8637.74229781226\\
0.39 8280.75926652225\\
0.395 7930.38115863161\\
0.4 7587.70870095927\\
0.405 7253.87440397576\\
0.41 6929.18343497213\\
0.415 6613.83232535911\\
0.42 6308.47072711763\\
0.425 6013.43857569102\\
0.43 5728.80986195373\\
0.435 5455.15886909077\\
0.44 5191.55055970602\\
0.445 4938.41614026117\\
0.45 4696.3143507546\\
0.455 4464.01006284199\\
0.46 4241.92879171642\\
0.465 4030.29224825574\\
0.47 3827.79528336227\\
0.475 3635.08052741638\\
0.48 3451.39699492463\\
0.485 3276.94277257903\\
0.49 3110.62718778146\\
0.495 2952.76019449803\\
0.5 2802.86914793216\\
};
\addplot [
color=red,
solid,
line width=3.0pt
]
table[row sep=crcr]{
0.01 120.485251052343\\
0.015 294.39944215877\\
0.02 553.215812560543\\
0.025 896.86414270765\\
0.03 1320.20296472896\\
0.035 1814.50942738593\\
0.04 2371.59137105316\\
0.045 2981.19490477423\\
0.05 3632.8511631575\\
0.055 4318.87899672681\\
0.06 5030.2438482813\\
0.065 5759.50296276871\\
0.07 6500.19978768681\\
0.075 7246.09791198797\\
0.08 7993.03976992883\\
0.085 8735.95123283668\\
0.09 9472.57042585678\\
0.095 10199.0243510491\\
0.1 10912.9785557975\\
0.105 11611.4470032359\\
0.11 12294.3200853638\\
0.115 12958.5202041012\\
0.12 13603.2303483007\\
0.125 14228.0309889982\\
0.13 14830.4905249417\\
0.135 15410.0897185394\\
0.14 15967.661567493\\
0.145 16499.6898680185\\
0.15 17007.8782705103\\
0.155 17490.8948044565\\
0.16 17947.8905215684\\
0.165 18379.3326298465\\
0.17 18782.2125753972\\
0.175 19157.5618729673\\
0.18 19506.3132435143\\
0.185 19826.3727104774\\
0.19 20117.6782446739\\
0.195 20378.3104420388\\
0.2 20610.3186727519\\
0.205 20810.3957617114\\
0.21 20980.9341582704\\
0.215 21120.2243934002\\
0.22 21228.6111773688\\
0.225 21303.9480820198\\
0.23 21348.5499220409\\
0.235 21360.8513120787\\
0.24 21340.9228434507\\
0.245 21289.5397146102\\
0.25 21206.2859656157\\
0.255 21091.0393474717\\
0.26 20944.9623597979\\
0.265 20767.9525703552\\
0.27 20562.2747288852\\
0.275 20325.6422106819\\
0.28 20061.5050326018\\
0.285 19770.4532686443\\
0.29 19453.1781767292\\
0.295 19111.5689239045\\
0.3 18746.9329262558\\
0.305 18361.2302944745\\
0.31 17956.5777509425\\
0.315 17532.0715393794\\
0.32 17093.0649544851\\
0.325 16639.7072030836\\
0.33 16174.8073085842\\
0.335 15699.9918897088\\
0.34 15217.6324340081\\
0.345 14728.6193569096\\
0.35 14236.6979540096\\
0.355 13742.3946152028\\
0.36 13248.8650365252\\
0.365 12756.8682631963\\
0.37 12268.2655839595\\
0.375 11785.0115308502\\
0.38 11308.2916414958\\
0.385 10839.5636950715\\
0.39 10380.3416805442\\
0.395 9930.63584519309\\
0.4 9491.70995394723\\
0.405 9064.87252081923\\
0.41 8650.47746400365\\
0.415 8248.74378995907\\
0.42 7860.3824604139\\
0.425 7485.74860922623\\
0.43 7124.88683256082\\
0.435 6778.39541531119\\
0.44 6445.17892868705\\
0.445 6125.65744439085\\
0.45 5820.39576686561\\
0.455 5527.94537297393\\
0.46 5248.71332513022\\
0.465 4982.86949391712\\
0.47 4728.8899411949\\
0.475 4487.41474487875\\
0.48 4257.52988811369\\
0.485 4039.38819377971\\
0.49 3831.69864626526\\
0.495 3634.73353141498\\
0.5 3447.90081522592\\
};
\tikzstyle{every node}=[font=\LARGE]
\end{axis}
\begin{scope}[x={(image.south east)},y={(image.north west)},shift={(-18pt,0)}]
\node at (0.62,0.8) {(d)};
\node at (-0.09,0.8) {\textcolor{white}{(a)}};
\node[anchor=south west,inner sep=0] at (0.51,0.03) {\raisebox{0pt}{\includegraphics[scale=0.05]{backward_mode1.png}}};
\draw[solid,red,line width=2.0pt] (0.32,1.24) -- (0.52,1.24);
\draw[solid,green,line width=2.0pt] (0.32,1.12) -- (0.52,1.12);
\draw[solid,blue,line width=2.0pt] (0.32,1) -- (0.52,1);
\draw[loosely dotted,red,line width=2.0pt] (0.72,1.24) -- (0.92,1.24);
\draw[loosely dotted,green,line width=2.0pt] (0.72,1.12) -- (0.92,1.12);
\draw[loosely dotted,blue,line width=2.0pt] (0.72,1) -- (0.92,1);
\node[scale=0.8] at (0.62,1.24) {$G$};
\node[scale=0.8] at (0.62,1.12) {$G_{\text{rp}}$};
\node[scale=0.8] at (0.62,1) {$G_{\text{es}}$};
\node[scale=0.8] at (0.22,1.3) {$g=$};
\node[scale=0.8] at (0.42,1.31) {5};
\node[scale=0.8] at (0.8,1.31) {50};
\node[scale=0.8] at (0.98,1.3) {nm};
\end{scope}
\end{tikzpicture}%
}
\vspace{-6mm}
\caption{(a) Gradient forces can be large despite low dispersion, (b-d) Narrow slots perform better than a stand-alone wire for a range of $\bar{a}$-values and (c) $G$ has a clear optimum in the  $(a,b)$-plane for the same mode as in (b) with $g=5 \, \text{nm}$.}
\vspace{-6mm}

\end{figure}

Next, we investigate the effect of $\bar{a}$ (fig.5b-d). As $\bar{a} \rightarrow 0$, there is no slot-enhancement. Then $G \rightarrow \tilde{G}$, regardless of all other parameters. Furthermore, the optical mode increasingly retreats into the widest beam. This implies $G \rightarrow 0$ when $\bar{a} \rightarrow \infty$, although this effect is more pronounced in wider slots.

In the forward case (fig.5b), $\bar{a}$ affects only the force distribution. The gain $G(\bar{a})$ has a maximum in narrow slots, but decreases monotonically otherwise. This confirms that small gaps are required for substantial SBS gain enhancement in vertical slot waveguides.

In the backward case (fig.5d), $G(\bar{a})$ always has a maximum because this phonon is forbidden in a stand-alone wire. However, the maximum increases by a factor $26$ when the slot is narrowed from $50 \, \text{nm}$ to $5 \, \text{nm}$. The gain is dominated by gradient forces regardless of $(\bar{a},g)$.

Last, we scan $(a,b)$ with $\bar{a} = a$ and $g$ fixed at $5 \, \text{nm}$. These parameters influence both the optical and mechanical mode. The $(a,b)$-optimum depends heavily on the slot size and on the mechanical mode. Nonetheless, fig.5c shows that there actually exists such an optimum. We find a maximum gain of $7.0 \times10^{4} \, \text{W}^{-1}\text{m}^{-1}$ for $(a,b)=(260,150) \, \text{nm}$.

\section{SBS in horizontal slot waveguides}

The horizontal slot (fig.1c-d) has the potential advantage of (1) the extra degree of freedom $\bar{b}$ and (2) smaller gaps. In such a slot, $g$ is not limited by the resolution of lithography techniques. As a result, SBS enhancement may be within reach of current technology. As long as $\bar{b}=b$, the horizontal slot waveguide is but a rotated version of the vertical one. Therefore we immediately explore the case $\bar{b} \neq b$. We calculate the forward and backward Brillouin spectrum for a horizontal slot waveguide with dimensions $(a,b,\bar{a},\bar{b},g)=(160,620,a,240,5) \, \text{nm}$.

In the forward case (fig.6a), the fundamental flexural mode couples most efficiently. This mode has negligible SBS gain in a stand-alone wire because of cancellations in the photon-phonon overlap. Indeed, the $u_{y}$ component has two nodes, while the $y$-component of the gradient force does not change sign. Owing to $b > \bar{b}$, the cancellations can be avoided by confining the optical mode between the nodes of $u_{y}$.

In the backward case (fig.6c), there are two modes with enhanced SBS gain. The first mode has a nearly uniform $u_{y}$ component. It is a rotated version of the mode we previously studied in fig.4-5d. The second mode is the fundamental flexural mode, but at the operating point $K \approx 2\beta$ in its dispersion diagram.

The gain increases by four orders of magnitude when $g$ drops from $250$ to $5 \, \text{nm}$ (fig.6b). This radical enhancement is superexponential in $g$ for gaps below $50 \, \text{nm}$. The forward (backward) gain approaches $\approx 1.3 \times10^{6} \, \text{W}^{-1}\text{m}^{-1}$  ($1.5 \times10^{5} \, \text{W}^{-1}\text{m}^{-1}$) as $g \rightarrow 0$. At $g = 70 \, \text{nm}$, an optical mode anti-crossing causes a dip in the SBS gain. However, $G(g)$ quickly recovers its original path as $g$ leaves the anti-cross region. We only show the total gain $G$ because $G_{\text{es}}$ is at least a factor $10^{5}$ ($10^{2}$) smaller than $G_{\text{rp}}$ across the entire sweep range in the forward (backward) case. Thus SBS by these modes is driven by gradient forces only, with a vanishing electrostrictive contribution.

Finally, we sweep $\bar{b}$ (fig.6d). In the forward case, $k_{\text{eff}}$ and $\mathbf{u}$ do not depend on $\bar{b}$. Then we explore purely the effect of the gradient force density $\mathbf{f_{\text{rp}}}(\bar{b})$ on the photon-phonon overlap $\langle \mathbf{f_{\text{rp}}}(\bar{b}), \mathbf{u}\rangle$. The coupling is optimal for $\bar{b}=240 \, \text{nm}$. For smaller $\bar{b}$, $G$ decreases because the slot-enhancement occurs only in a small region. For larger $\bar{b}$, $G$ decreases because the optical mode is no longer confined between the nodes of $u_{y}$. In the backward case, the operating point $K \approx 2\beta$ changes as $n_{p}$ depends on $\bar{b}$. This propagating phonon is less sensitive to $\bar{b}$ because of its nearly uniform $u_{y}$ component.
\begin{figure}

\centering
\label{fig:f}
\subfigure{\begin{tikzpicture}[scale = 0.44*\columnwidth/(3.5in)]
\pgfplotsset{try min ticks=3}
\pgfplotsset{max space between ticks=50pt}
\begin{axis}[%
width=3.5in,
height=2.33in,
xmin=1,
xmax=13,
xlabel={Frequency (GHz)},
ymin=0,
ymax=4.5,
ylabel={Gain ($\text{W}^{-1}\text{m}^{-1}$)},
legend style={fill=none,draw=none,legend cell align=left,at={(0.5,-0.41)},anchor=north, legend columns=-1},
xtick = {2,4,6,8,10,12}
]
\addplot [
color=black,
solid,
line width=2.0pt,
forget plot
]
table[row sep=crcr]{
3.16114747521379 0\\
3.16114747521379 3.9056184484259\\
};
\addplot [
color=black,
solid,
line width=2.0pt,
forget plot
]
table[row sep=crcr]{
11.0942239977886 0\\
11.0942239977886 0.206730610842735\\
};
\addplot [
color=black,
solid,
line width=2.0pt,
forget plot
]
table[row sep=crcr]{
0 0\\
0 0\\
};
\addplot [
color=black,
solid,
line width=2.0pt,
forget plot
]
table[row sep=crcr]{
0 0\\
0 0\\
};
\addplot [
color=black,
solid,
line width=2.0pt,
forget plot
]
table[row sep=crcr]{
0 0\\
0 0\\
};
\addplot [
color=black,
solid,
line width=2.0pt,
forget plot
]
table[row sep=crcr]{
0 0\\
0 0\\
};
\addplot [
color=black,
solid,
line width=2.0pt,
forget plot
]
table[row sep=crcr]{
0 0\\
0 0\\
};
\addplot [
color=black,
solid,
line width=2.0pt,
forget plot
]
table[row sep=crcr]{
0 0\\
0 0\\
};
\addplot [
color=black,
solid,
line width=2.0pt,
forget plot
]
table[row sep=crcr]{
0 0\\
0 0\\
};
\addplot [
color=black,
solid,
line width=2.0pt,
forget plot
]
table[row sep=crcr]{
0 0\\
0 0\\
};
\addplot [
color=blue,
line width=3.0pt,
mark size=5.0pt,
only marks,
mark=x,
mark options={solid}
]
table[row sep=crcr]{
0 0\\
0 0\\
0 0\\
0 0\\
3.16114747521379 2.18152800727311e-005\\
0 0\\
0 0\\
0 0\\
0 0\\
11.0942239977886 0.000969337633328789\\
0 0\\
0 0\\
0 0\\
0 0\\
0 0\\
0 0\\
0 0\\
0 0\\
0 0\\
0 0\\
0 0\\
0 0\\
0 0\\
0 0\\
0 0\\
0 0\\
0 0\\
0 0\\
0 0\\
0 0\\
};
\addlegendentry{$G_{\text{es}}$ \, \,};
\addplot [
color=green,
line width=3.0pt,
mark size=7.0pt,
only marks,
mark=+,
mark options={solid}
]
table[row sep=crcr]{
0 0\\
0 0\\
0 0\\
0 0\\
3.16114747521379 3.9056184484259\\
0 0\\
0 0\\
0 0\\
0 0\\
11.0942239977886 0.179387989091315\\
0 0\\
0 0\\
0 0\\
0 0\\
0 0\\
0 0\\
0 0\\
0 0\\
0 0\\
0 0\\
0 0\\
0 0\\
0 0\\
0 0\\
0 0\\
0 0\\
0 0\\
0 0\\
0 0\\
0 0\\
};
\addlegendentry{$G_{\text{rp}}$ \, \,};
\addplot [
color=red,
line width=3.0pt,
mark size=3.7pt,
only marks,
mark=*,
mark options={solid}
]
table[row sep=crcr]{
0 0\\
0 0\\
0 0\\
0 0\\
3.16114747521379 3.88717926044941\\
0 0\\
0 0\\
0 0\\
0 0\\
11.0942239977886 0.206730610842735\\
0 0\\
0 0\\
0 0\\
0 0\\
0 0\\
0 0\\
0 0\\
0 0\\
0 0\\
0 0\\
0 0\\
0 0\\
0 0\\
0 0\\
0 0\\
0 0\\
0 0\\
0 0\\
0 0\\
0 0\\
};
\addlegendentry{$G$};
\tikzstyle{every node}=[font=\LARGE]
\end{axis}
\begin{scope}[x={(image.south east)},y={(image.north west)},shift={(-18pt,0)}]
\node[scale=0.8] at (0.18,0.77) {$\times 10^{5}$};
\node at (0.62,0.8) {(a)};
\node at (1.2,0.8) {\textcolor{white}{(a)}};
\node[anchor=south west,inner sep=0] at (0.31,0.49) {\raisebox{0pt}{\includegraphics[scale=0.06]{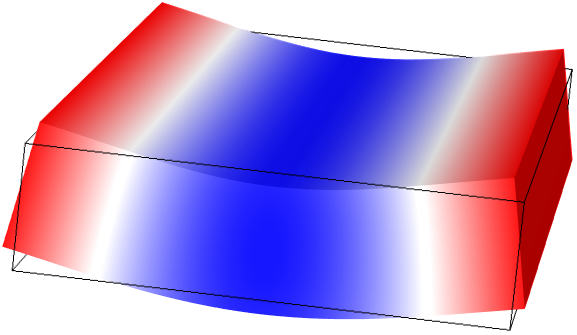}}};
\node[anchor=south west,inner sep=0] at (0.8,0.07) {\raisebox{0pt}{\includegraphics[scale=0.06]{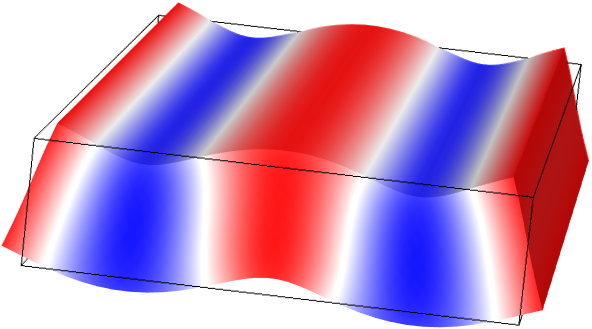}}};
\node[scale = 0.9] at (0.88,0.63) {\small{Forward}};
\end{scope}
\end{tikzpicture}%
}
\subfigure{\begin{tikzpicture}[scale = 0.44*\columnwidth/(3.5in)]
\pgfplotsset{try min ticks=3}
\pgfplotsset{max space between ticks=50pt}
\begin{axis}[%
width=3.5in,
height=2.33in,
xmin=0,
xmax=0.25,
xlabel={$g$ ($\mu$m)},
ymode=log,
ymin=10,
ymax=1000000,
yminorticks=false,
ylabel={Gain ($\text{W}^{-1}\text{m}^{-1}$)},
legend style={fill=none,draw=none,legend cell align=left,at={(0.5,-0.41)},anchor=north, legend columns=-1},
xtick = {0,0.05,0.1,0.15,0.2,0.25},
xticklabel style={/pgf/number format/fixed},
ytick = {100,10000,1000000},
extra y ticks = {1000,100000},
extra y tick labels={}
]
\addplot [
color=red,
solid,
line width=3.0pt
]
table[row sep=crcr]{
0.0025 685761.068156646\\
0.005 396143.474187155\\
0.0075 253374.916115238\\
0.01 173846.856598546\\
0.0125 125557.683950459\\
0.015 94317.8767764763\\
0.0175 73092.7539844482\\
0.02 58091.8219224591\\
0.0225 47136.906237912\\
0.025 38923.4587587488\\
0.0275 32622.0559329265\\
0.03 27691.2415406624\\
0.0325 23766.876592355\\
0.035 20597.9398918932\\
0.0375 18007.1781601803\\
0.04 15864.0409927641\\
0.0425 14074.2898938506\\
0.045 12567.0779404451\\
0.0475 11289.5898743469\\
0.05 10200.9928913597\\
0.0525 9270.04894865838\\
0.055 8476.46572009508\\
0.0575 7802.45822812921\\
0.06 7244.88250935434\\
0.0625 6817.81973715747\\
0.065 6578.68000208151\\
0.0675 6810.88626391228\\
0.07 1741.46607557274\\
0.0725 2291.55905616496\\
0.075 3130.17171288184\\
0.0775 3203.93434570728\\
0.08 3115.12783533681\\
0.0825 2977.17431467206\\
0.085 2824.63312639913\\
0.0875 2671.99034984888\\
0.09 2523.30572302418\\
0.0925 2382.56982464764\\
0.095 2248.83109602499\\
0.0975 2123.84408343526\\
0.1 2006.91688543349\\
0.1025 1896.79468874605\\
0.105 1794.53498528589\\
0.1075 1698.95652288272\\
0.11 1609.32376248111\\
0.1125 1525.69727422295\\
0.115 1447.5192904702\\
0.1175 1374.01799362687\\
0.12 1305.45992956213\\
0.1225 1240.73655748614\\
0.125 1180.14481663362\\
0.1275 1123.46096813977\\
0.13 1069.65089576066\\
0.1325 1019.060794684\\
0.135 971.560061347554\\
0.1375 926.103909434597\\
0.14 884.212138700038\\
0.1425 844.10658005639\\
0.145 806.287433786511\\
0.1475 769.835296821978\\
0.15 736.218471765752\\
0.1525 704.023787824311\\
0.155 673.506127108578\\
0.1575 644.334409574609\\
0.16 616.67991354269\\
0.1625 590.380459230677\\
0.165 565.310057479462\\
0.1675 541.538857333161\\
0.17 518.852213249239\\
0.1725 497.316608335463\\
0.175 476.420744482769\\
0.1775 456.748352800151\\
0.18 437.858126821703\\
0.1825 419.924915730706\\
0.185 402.575765150194\\
0.1875 386.095975542991\\
0.19 370.298416627539\\
0.1925 355.082646071784\\
0.195 340.661504206146\\
0.1975 326.765441886966\\
0.2 313.505639492654\\
0.2025 300.693742884633\\
0.205 288.534550955479\\
0.2075 276.673192044511\\
0.21 265.337629284901\\
0.2125 254.439794365034\\
0.215 243.92321762569\\
0.2175 233.971243811911\\
0.22 224.304927046728\\
0.2225 215.045673946023\\
0.225 206.033631331493\\
0.2275 197.430237382483\\
0.23 189.108330819734\\
0.2325 181.129785786974\\
0.235 173.494708664922\\
0.2375 166.136603459487\\
0.24 159.001417341888\\
0.2425 152.044112890494\\
0.245 145.416303490216\\
0.2475 139.072720835514\\
0.25 132.91907568023\\
};
\addplot [
color=red,
loosely dotted,
line width=3.0pt
]
table[row sep=crcr]{
0.0025 96820.1982484367\\
0.005 67214.4295788935\\
0.0075 49554.9402691761\\
0.01 37988.7939170015\\
0.0125 29916.8635261575\\
0.015 24035.4103502099\\
0.0175 19609.5047274021\\
0.02 16194.0688779834\\
0.0225 13504.9075965214\\
0.025 11352.7719889053\\
0.0275 9607.49093068984\\
0.03 8175.37528280058\\
0.0325 6989.54293820268\\
0.035 5998.3963727023\\
0.0375 5164.59994607655\\
0.04 4458.27381550835\\
0.0425 3856.32026076638\\
0.045 3340.13168644136\\
0.0475 2895.06074229601\\
0.05 2508.47773040804\\
0.0525 2171.04033777281\\
0.055 1872.49806730107\\
0.0575 1605.01051938452\\
0.06 1359.16377644773\\
0.0625 1120.43125620924\\
0.065 863.620222835284\\
0.0675 504.665358717706\\
0.07 3149.32905096488\\
0.0725 1956.25508771459\\
0.075 1197.48442739156\\
0.0775 932.554622879107\\
0.08 775.097927043454\\
0.0825 661.776572487622\\
0.085 573.018099333007\\
0.0875 499.848510912074\\
0.09 438.44670562767\\
0.0925 385.724396455406\\
0.095 340.350016730112\\
0.0975 300.757877812828\\
0.1 266.110600898858\\
0.1025 235.982934430716\\
0.105 208.919315778672\\
0.1075 185.280544607868\\
0.11 164.341993718254\\
0.1125 145.814089275549\\
0.115 129.394390232989\\
0.1175 114.798968075243\\
0.12 101.842806760493\\
0.1225 90.3451131075276\\
0.125 80.1239167484808\\
0.1275 70.9972936432377\\
0.13 62.9105010815601\\
0.1325 55.715191878231\\
0.135 49.3035383491945\\
0.1375 43.6636576595973\\
0.14 38.5198031791702\\
0.1425 34.0106390728113\\
0.145 29.9930756284091\\
0.1475 26.4816105957161\\
0.15 23.264454933385\\
0.1525 20.4549828437432\\
0.155 17.9758206224874\\
0.1575 15.7553143246748\\
0.16 13.7883590637385\\
0.1625 12.0573469221859\\
0.165 10.5233610076178\\
0.1675 9.16730353154536\\
0.17 7.97106715064096\\
0.1725 6.91467847068273\\
0.175 5.99197733152243\\
0.1775 5.17658895016143\\
0.18 4.46228804174972\\
0.1825 3.83608869098644\\
0.185 3.2892040939259\\
0.1875 2.81168674943413\\
0.19 2.39511619278128\\
0.1925 2.0444974351116\\
0.195 1.72195646756058\\
0.1975 1.45222758033215\\
0.2 1.2192971432147\\
0.2025 1.0201308594592\\
0.205 0.848357312380959\\
0.2075 0.703774656049839\\
0.21 0.580384054934045\\
0.2125 0.475574919382775\\
0.215 0.389407199773163\\
0.2175 0.313980648983176\\
0.22 0.252436296950852\\
0.2225 0.202211934671051\\
0.225 0.160592865322753\\
0.2275 0.126681193943385\\
0.23 0.0990681864036655\\
0.2325 0.0768728958535685\\
0.235 0.0591219690345652\\
0.2375 0.0451611389751964\\
0.24 0.0343332832050054\\
0.2425 0.0262422853526623\\
0.245 0.0198477192058494\\
0.2475 0.0149918781155845\\
0.25 0.0115470305368659\\
};
\tikzstyle{every node}=[font=\LARGE]
\end{axis}
\begin{scope}[x={(image.south east)},y={(image.north west)},shift={(-18pt,0)}]
\node at (0.5,-0.38) {};
\node at (0.62,0.8) {(b)};
\node[anchor=south west,inner sep=0] at (0.28,-0.43) {\raisebox{0pt}{\includegraphics[scale=0.05]{forward_mode1_hor.png}}};
\node[anchor=south west,inner sep=0] at (0.87,-0.43) {\raisebox{0pt}{\includegraphics[scale=0.05]{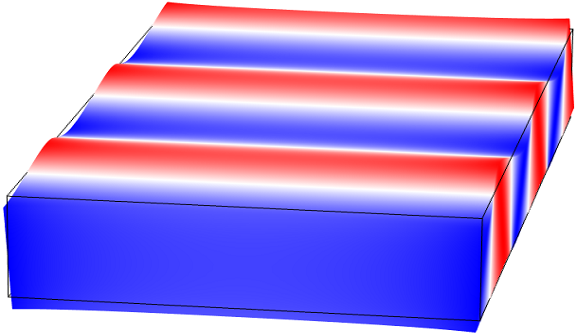}}};
\draw[solid,red,line width=2.0pt] (0.09,-0.36) -- (0.26,-0.36);
\draw[loosely dotted,red,line width=2.0pt] (0.68,-0.36) -- (0.85,-0.36);
\draw [->, thick] (0.4,0.4) to [out=90, in=230] (0.54,0.55);
\node[scale=0.8] at (0.79,0.6) {Anti-crossing};
\end{scope}
\end{tikzpicture}%
}
\subfigure{\begin{tikzpicture}[scale = 0.44*\columnwidth/(3.5in)]
\pgfplotsset{try min ticks=3}
\pgfplotsset{max space between ticks=50pt}
\begin{axis}[%
width=3.5in,
height=2.33in,
xmin=3,
xmax=15,
xlabel={Frequency (GHz)},
ymin=0,
ymax=0.7,
ylabel={Gain ($\text{W}^{-1}\text{m}^{-1}$)},
legend style={fill=none,draw=none,legend cell align=left,at={(0.5,-0.41)},anchor=north, legend columns=-1},
xtick = {4,6,8,10,12,14},
ytick = {0,0.3,0.6}
]
\addplot [
color=black,
solid,
line width=2.0pt,
forget plot
]
table[row sep=crcr]{
8.14963176588702 0\\
8.14963176588702 0.658128064017505\\
};
\addplot [
color=black,
solid,
line width=2.0pt,
forget plot
]
table[row sep=crcr]{
10.3618418716802 0\\
10.3618418716802 0.518509259018949\\
};
\addplot [
color=black,
solid,
line width=2.0pt,
forget plot
]
table[row sep=crcr]{
21.843964821274 0\\
21.843964821274 0.106081819582332\\
};
\addplot [
color=black,
solid,
line width=2.0pt,
forget plot
]
table[row sep=crcr]{
0 0\\
0 0.602039891312147\\
};
\addplot [
color=black,
solid,
line width=2.0pt,
forget plot
]
table[row sep=crcr]{
0 0\\
0 0.602039891312147\\
};
\addplot [
color=black,
solid,
line width=2.0pt,
forget plot
]
table[row sep=crcr]{
0 0\\
0 0.602039891312147\\
};
\addplot [
color=black,
solid,
line width=2.0pt,
forget plot
]
table[row sep=crcr]{
0 0\\
0 0.602039891312147\\
};
\addplot [
color=black,
solid,
line width=2.0pt,
forget plot
]
table[row sep=crcr]{
0 0\\
0 0.602039891312147\\
};
\addplot [
color=black,
solid,
line width=2.0pt,
forget plot
]
table[row sep=crcr]{
0 0\\
0 0.602039891312147\\
};
\addplot [
color=black,
solid,
line width=2.0pt,
forget plot
]
table[row sep=crcr]{
0 0\\
0 0.602039891312147\\
};
\addplot [
color=blue,
line width=3.0pt,
mark size=5.0pt,
only marks,
mark=x,
mark options={solid}
]
table[row sep=crcr]{
8.14963176588702 0.0012488187198816\\
0 0\\
10.3618418716802 0.00107452321698433\\
0 0\\
0 0\\
0 0\\
0 0\\
0 0\\
0 0\\
0 0\\
0 0\\
0 0\\
0 0\\
0 0\\
0 0\\
21.843964821274 0.00423420144915773\\
0 0\\
0 0\\
0 0\\
0 0\\
0 0\\
0 0\\
0 0\\
0 0\\
0 0\\
0 0\\
0 0\\
0 0\\
0 0\\
0 0\\
};
\addplot [
color=green,
line width=3.0pt,
mark size=5.0pt,
only marks,
mark=+,
mark options={solid}
]
table[row sep=crcr]{
8.14963176588702 0.602039891312147\\
0 0\\
10.3618418716802 0.472375722032735\\
0 0\\
0 0\\
0 0\\
0 0\\
0 0\\
0 0\\
0 0\\
0 0\\
0 0\\
0 0\\
0 0\\
0 0\\
21.843964821274 0.0679286741362285\\
0 0\\
0 0\\
0 0\\
0 0\\
0 0\\
0 0\\
0 0\\
0 0\\
0 0\\
0 0\\
0 0\\
0 0\\
0 0\\
0 0\\
};
\addplot [
color=red,
line width=3.0pt,
mark size=3.7pt,
only marks,
mark=*,
mark options={solid}
]
table[row sep=crcr]{
8.14963176588702 0.658128064017505\\
0 0\\
10.3618418716802 0.518509259018949\\
0 0\\
0 0\\
0 0\\
0 0\\
0 0\\
0 0\\
0 0\\
0 0\\
0 0\\
0 0\\
0 0\\
0 0\\
21.843964821274 0.106081819582332\\
0 0\\
0 0\\
0 0\\
0 0\\
0 0\\
0 0\\
0 0\\
0 0\\
0 0\\
0 0\\
0 0\\
0 0\\
0 0\\
0 0\\
};
\tikzstyle{every node}=[font=\LARGE]
\end{axis}
\begin{scope}[x={(image.south east)},y={(image.north west)},shift={(-18pt,0)}]
\node[scale=0.8] at (0.18,0.77) {$\times 10^{5}$};
\node at (0.62,0.8) {(c)};
\node at (1.2,0.8) {\textcolor{white}{(a)}};
\node[anchor=south west,inner sep=0] at (0.18,0.45) {\raisebox{0pt}{\includegraphics[scale=0.06]{backward_mode1_hor.png}}};
\node[anchor=south west,inner sep=0] at (0.76,0.31) {\raisebox{0pt}{\includegraphics[scale=0.06]{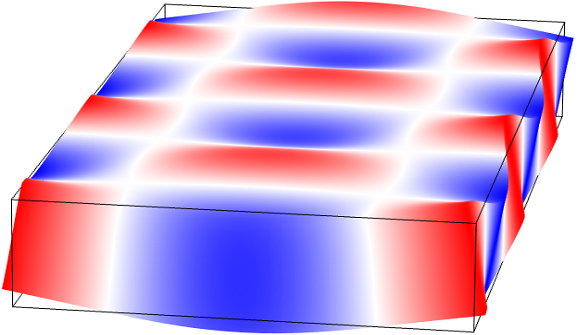}}};
\node[scale = 0.9] at (0.87,0.63) {\small{Backward}};
\end{scope}
\end{tikzpicture}%
}
\subfigure{\begin{tikzpicture}[scale = 0.44*\columnwidth/(3.5in)]
\pgfplotsset{try min ticks=3}
\pgfplotsset{max space between ticks=50pt}
\begin{axis}[%
width=3.5in,
height=2.33in,
xmin=0.1,
xmax=0.8,
xlabel={$\bar{b}$ ($\mu$m)},
ymode=log,
ymin=3000,
ymax=1000000,
yminorticks=false,
ylabel={Gain ($\text{W}^{-1}\text{m}^{-1}$)},
legend style={fill=none,draw=none,legend cell align=left,at={(0.5,-0.41)},anchor=north, legend columns=-1},
xticklabel style={/pgf/number format/fixed},
ytick = {1e4,1e5,1e6},
extra y ticks = {100,10000},
extra y tick labels={}
]
\addplot [
color=red,
solid,
line width=3.0pt
]
table[row sep=crcr]{
0.1 108910.573535186\\
0.11 145598.504435185\\
0.12 182479.247745881\\
0.13 217917.989108667\\
0.14 250703.043474798\\
0.15 280100.133047145\\
0.16 305762.324809461\\
0.17 327606.279396686\\
0.18 345749.53440706\\
0.19 360382.244878929\\
0.2 371755.017210369\\
0.21 380155.201035675\\
0.22 385859.427050355\\
0.23 389127.361063244\\
0.24 390224.377795632\\
0.25 389395.331782451\\
0.26 386879.690730088\\
0.27 382876.805416296\\
0.28 377603.667042975\\
0.29 371220.752649545\\
0.3 363912.626416717\\
0.31 355818.625978748\\
0.32 347068.74121329\\
0.33 337781.144609341\\
0.34 328066.070831601\\
0.35 318031.428825875\\
0.36 307745.415703386\\
0.37 297305.962556619\\
0.38 286776.816511971\\
0.39 276219.361881121\\
0.4 265685.497686784\\
0.41 255219.515819501\\
0.42 244867.369883698\\
0.43 234672.599983965\\
0.44 224672.785869569\\
0.45 214899.986290255\\
0.46 205378.620611071\\
0.47 196126.015814148\\
0.48 187154.582014566\\
0.49 178481.823934557\\
0.5 170112.86385222\\
0.51 162062.715160369\\
0.52 154336.533726708\\
0.53 146947.595612721\\
0.54 139893.455217844\\
0.55 133151.593388976\\
0.56 126733.318853697\\
0.57 120625.969689032\\
0.58 114831.802514615\\
0.59 109352.188299617\\
0.6 104260.614782575\\
0.61 99782.4083530878\\
0.62 96133.8294625227\\
0.63 92839.6931084282\\
0.64 89135.0142004857\\
0.65 85033.2416492269\\
0.66 80554.3704192916\\
0.67 75594.8229447561\\
0.68 70062.5682495145\\
0.69 63873.2138017902\\
0.7 56908.2419964193\\
0.71 49206.6678058411\\
0.72 40884.5589522212\\
0.73 32272.4611945583\\
0.74 23931.4485256641\\
0.75 16473.1641296625\\
0.76 10442.7182344164\\
0.77 6053.7149919009\\
0.78 3179.65761810614\\
0.79 1469.36528697734\\
0.8 558.500436739206\\
};
\addplot [
color=red,
loosely dotted,
line width=3.0pt,
]
table[row sep=crcr]{
0.1 7570.26149919596\\
0.11 15002.1055199075\\
0.12 23196.834171897\\
0.13 31162.3535225837\\
0.14 38335.8089608811\\
0.15 44486.9437684263\\
0.16 49591.7101945546\\
0.17 53734.8465999057\\
0.18 57042.9061444723\\
0.19 59650.5549589476\\
0.2 61681.7626235141\\
0.21 63247.0629744353\\
0.22 64437.265209666\\
0.23 65324.6633976951\\
0.24 65969.296317748\\
0.25 66420.7633307857\\
0.26 66719.7848377467\\
0.27 66894.5986313562\\
0.28 66975.0784540068\\
0.29 66981.8669106061\\
0.3 66933.5641435816\\
0.31 66844.1750732861\\
0.32 66725.427950693\\
0.33 66586.8247999141\\
0.34 66432.5661106919\\
0.35 66266.2130114212\\
0.36 66100.8455554515\\
0.37 65930.6957753015\\
0.38 65764.2817160243\\
0.39 65603.7976225498\\
0.4 65451.7262005921\\
0.41 65309.9362047731\\
0.42 65181.8459519917\\
0.43 65066.7547131442\\
0.44 64963.4643669621\\
0.45 64873.5937536349\\
0.46 64797.6262065963\\
0.47 64737.7589040671\\
0.48 64693.8647220801\\
0.49 64667.6232815456\\
0.5 64659.6174124478\\
0.51 64668.7538321151\\
0.52 64694.8195005829\\
0.53 64736.2981565143\\
0.54 64792.530118748\\
0.55 64865.0674114908\\
0.56 64950.9763361267\\
0.57 65047.1127720997\\
0.58 65148.3860179271\\
0.59 65247.6322876493\\
0.6 65325.8499644677\\
0.61 65339.6353639249\\
0.62 65255.4446343057\\
0.63 65126.1178902398\\
0.64 64986.2166007608\\
0.65 64766.1031993968\\
0.66 64392.6221858275\\
0.67 63791.4736712993\\
0.68 62864.9752475971\\
0.69 61486.382310163\\
0.7 59507.7289959618\\
0.71 56752.9495184588\\
0.72 53070.2931645057\\
0.73 48387.8911454061\\
0.74 42800.887079723\\
0.75 36636.2835136336\\
0.76 30386.5829800423\\
0.77 24582.7423835223\\
0.78 19569.3683746685\\
0.79 15487.2016020743\\
0.8 12298.599816257\\
};
\tikzstyle{every node}=[font=\LARGE]
\end{axis}

\begin{scope}[x={(image.south east)},y={(image.north west)},shift={(-18pt,0)}]

\node at (0.62,0.8) {(d)};
\end{scope}
\end{tikzpicture}%
}
\vspace{-4mm}
\caption{ (a-b-c) Both forward and backward SBS is very efficient in narrow horizontal slots and (d) the flexural mode is sensitive to $\bar{b}$. The color of the modes indicates the sign of $u_{y}$ (red: $+$, blue: $-$).}
\vspace{-7mm}
\end{figure}
\section{Conclusion}
To conclude, we found that strong gradient forces improve the efficiency of Brillouin scattering in narrow silicon slot waveguides. However, appreciable enhancement compared to a stand-alone wire is currently only accessible in horizontal slots. In such slots, we expect very efficient SBS because (1) small gaps should be technologically feasible and (2) the fundamental mechanical flexural mode can be excited. The suspension of long silicon beams remains the most important hurdle towards testing these predictions. A practical device may consist of a disconnected series of such waveguides as in \cite{Shin2013b}.
\section*{Supplementary information}
We use isotropic elasticity coefficients $(c_{11},c_{12},c_{44})=(217,85,66)\,\text{GPa}$ for easy comparison with \cite{Rakich2012, Qiu2012}. Silicon is mechanically anisotropic, so in a more accurate calculation the coefficients $(c_{11},c_{12},c_{44})=(166,64,80)\,\text{GPa}$ should be used for a guide along a $\langle 100 \rangle$ crystal axis \cite{Hopcroft2010b}. Further, we use the photoelastic coefficients $(p_{11},p_{12},p_{44}) = (-0.094,0.017,-0.051)$ \cite{Biegelsen1974}, which is also valid in case the guide is aligned along a $\langle 100 \rangle$ axis. We perform our calculations using the weak-form \cite{Dasgupta2011} {\textsf{\small{COMSOL}}} module with the {\textsf{\small{MATLAB Livelink}}}.
\section*{Acknowledgement}
R.V.L. acknowledges the Agency for Innovation by Science and Technology in Flanders (IWT) for a PhD grant. This work was partially funded under the FP7-ERC-\textit{InSpectra} programme. R.V.L. thanks Thomas Van Vaerenbergh for helpful discussions.
%


\end{document}